\documentclass[sigconf, anonymous=false, review=false, screen, 9pt]{acmart} 

\usepackage{xcolor}
 \usepackage{url}
\usepackage{xurl}
\usepackage{bm}
\usepackage{booktabs}
\usepackage[ruled]{algorithm2e}
\usepackage{algpseudocode}
\usepackage{enumerate}
\usepackage[english]{babel}
\usepackage{blindtext}
\usepackage[caption=false,font=footnotesize,subrefformat=parens]{subfig}
\usepackage{amsmath}

\usepackage{amssymb}
\usepackage{xspace}
\usepackage{multirow}
\usepackage{threeparttable}
\usepackage{float}
\usepackage{flafter}
\usepackage{comment}
\usepackage{mathtools}
\usepackage[skip=12pt]{caption}
\usepackage{graphicx}
\usepackage{gensymb}
\usepackage{amsfonts}
\usepackage{braket}
\usepackage{graphics, transparent}
\usepackage{svg}
\usepackage{graphicx}
\usepackage[qm]{qcircuit}
\usepackage{bbm}
\usepackage{cleveref}
\usepackage{extarrows}

\usepackage{stfloats}

\newcommand{\one}{\mathbbm{1}}
\newcommand{\change}{\par \noindent}

\def\fig{Fig.\xspace}

\def\tab{Tab.\xspace}

\def\ie{{\textit{i.e.}\xspace}} 
\def\eg{{\textit{e.g.}\xspace}}
 
\def\etal{{\textit{et al.}\xspace~}}
\def\etc{{\textit{etc}\xspace}}

\graphicspath{{./figures/}}
\graphicspath{{./fig_mobihoc/}}
\DeclareGraphicsExtensions{.pdf,.jpeg,.png}

\usepackage{fp}

\newlength\maxlentime

\ifodd 1
\newcommand{\com}[1]{\textbf{\color{red}(COMMENT: #1)}} 
\newcommand{\todo}[1]{\textbf{{\color{orange}(TODO: #1)}}}
\newcommand{\unused}[1]{}
\else

\newcommand{\com}[1]{}
\newcommand{\todo}[1]{}
\newcommand{\unused}[1]{}
\fi

\newcommand{\ppt}{\frac{\partial}{\partial t}}

\renewcommand\footnotetextcopyrightpermission[1]{} 
\setcopyright{none}

\acmDOI{}

\acmISBN{}

\acmConference[]{}
\copyrightyear{}
\acmSubmissionID{}

\acmPrice{}

\def\sysname{\textsc{QuERLoc}\xspace}
\def\rangname{\textsc{QuER}\xspace}
\begin{document}

\title[\sysname]{\sysname: Towards Next-Generation Localization with Quantum-Enhanced Ranging}


\begin{anonsuppress}
	\author{Entong He}
	\email{entong_he@connect.hku.hk}
	\affiliation{%
		\institution{University of Hong Kong}
		\city{Hong Kong SAR}
		\country{China}
	}

        \author{Yuxiang Yang}
	\email{yuxiang@cs.hku.hk}
	\affiliation{%
		\institution{University of Hong Kong}
		\city{Hong Kong SAR}
		\country{China}
	}
	
	\author{Chenshu Wu}
	\email{chenshu@cs.hku.hk}
	\affiliation{%
		\institution{University of Hong Kong}
		\city{Hong Kong SAR}
		\country{China}
	}
\end{anonsuppress}

\renewcommand{\shortauthors}{\sysname}

\begin{abstract}


Remarkable advances have been achieved in localization techniques in past decades, rendering it one of the most important technologies indispensable to our daily lives. In this paper, we investigate a novel localization approach for future computing by presenting \sysname, the first study on localization using quantum-enhanced ranging. By fine-tuning the evolution of an entangled quantum probe, quantum ranging can output the information integrated in the probe as a specific mapping of distance-related parameters. \sysname is inspired by this unique property to measure a special combination of distances between a target sensor and multiple anchors within one single physical measurement. Leveraging this capability, \sysname settles two drawbacks of classical localization approaches: \textbf{\textit{(i)}} the target-anchor distances must be measured individually and sequentially, and \textbf{\textit{(ii)}} the resulting optimization problems are non-convex and are sensitive to noise. We first present the theoretical formulation of preparing the probing quantum state and controlling its dynamic to induce a convexified localization problem, and then solve it efficiently via optimization. We conduct extensive numerical analysis of \sysname under various settings. The results show that \sysname consistently outperforms classical approaches in accuracy and closely follows the theoretical lowerbound, while maintaining low time complexity. It achieves a minimum reduction of 73\% in RMSE and 97.6\% in time consumption compared to baselines. By introducing range-based quantum localization to the mobile computing community and showing its superior performance, \sysname sheds light on next-generation localization technologies and opens up new directions for future research.


\end{abstract}

\maketitle

\section{Introduction}
\label{sec:intro}

Decades of efforts has been devoted into positioning techniques including the Global Positioning System (GPS) and various indoor localization systems \cite{indoor_loc_without_pain, Zee}, which have become an indispensable part of our lives. 
Range-based localization schemes locate sensors based on information of Euclidean distances between sensors and neighbouring anchor nodes, which is collected by their casual association with the received signals at sensors. They have been applied to different technologies, such as GPS, WiFi, mmWave/UWB radars and ultrasound, \etc. Typical ranging models to be exploited are the Time of Arrival (ToA), Time Difference of Arrival (TDoA), Angular of Arrival (AoA), and Received Signal Strength Indicator (RSSI) \cite{Sensor_Loc}. 
\par Apparently, the localization performance depends heavily on the ranging accuracy. In general, factors affecting the ranging accuracy of the existing methods is multiple-fold, including signal bandwidth, carrier frequency, and environmental effects \cite{Sensor_Loc}. For instance, accuracy of GPS degrades drastically in presence of blockings due to the deterioration of distance measurement, and WiFi localization suffers from low accuracy due to limited bandwidth \cite{Energy_Efficient_GPS}, synchronization offsets and multipath effect \cite{SpotFi}. 
Given that the signals are inaccurate, the distance estimates from the signals are consequently perturbed. 
In the classical picture, distances are measured individually (or in pairs) and sequentially. This could lead to either underutilization of available anchors or accumulated disturbances in the ranging model with the growth in the quantity of distance rangings. These often result in degraded solution quality, primarily due to the lack of correlations between each distance ranging. For instance, issues such as ill-conditioned least squares \cite{Least_Square} and large geometric dilution of precision (GDOP) affect trilateration-based localization. Additionally, significant deviations in the optimal solutions occur when maximum likelihood estimation (MLE) or error minimization \cite{SDP_localize, SOCP_Loc, SDP_TDoA_Loc} techniques are applied to location estimation. To make matters worse, range-based localization with classical ranging models is frequently reduced to an instance of non-convex optimization, which is NP-hard \cite{Handbook_of_SDP}. Although various techniques can be applied to convexify the objective, they could commonly cause problems such as lack of scalability and existence of optimality gap \cite{Handbook_of_SDP}. Despite the significant progress in enhancing ranging and achieving robust localization \cite{SpotFi, indoor_loc_without_pain, Noise_Tolerant_Loc, SDP_localize, MDS_MAP}, their accuracy remains inherently and physically limited within the classical framework.

\begin{figure}[t]
    \centering
    \includegraphics[width=\linewidth]{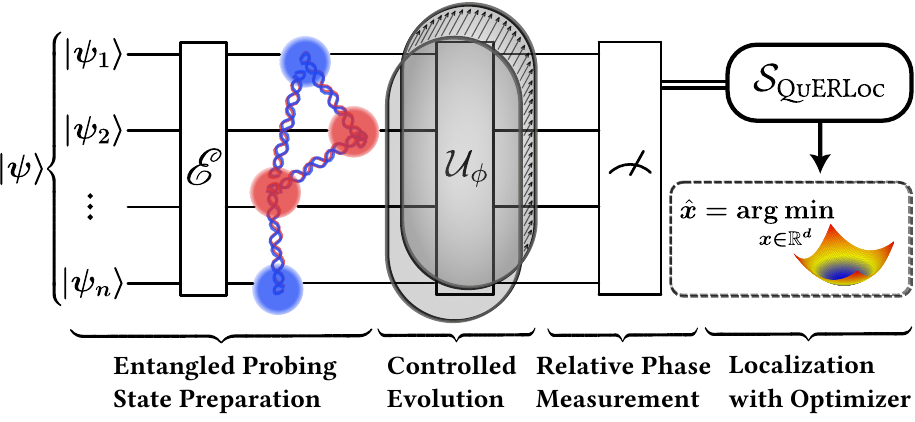}
    \caption{Workflow of \sysname}
    \label{fig:QuERLoc_workflow}
\end{figure}

\par In this paper, we foresee an emerging opportunity in quantum-based ranging to break such limits for next-generation localization systems. We propose \sysname, a localization approach assisted by \textit{Quantum-Enhanced Ranging (\rangname)}, which aims to concurrently tackle the co-existence of Gaussian noise accumulation from iterative ranging process, and the non-convexity objective arise from solving error minimization problem \cite{SDP_localize}. By introducing quantum metrology \cite{Beat_quantum_limit, quantum_metrology, AdvQuantumMetrology} and quantum control theory \cite{Quantum_Control, Quantum_Control_Bound_State} and harnessing entangled probing states, \rangname measures the Euclidean distances between the target and anchor nodes in a \textit{correlated} manner rather than an \textit{isolated} one. Schematic diagram is shown in \fig \ref{fig:QuERLoc_workflow}. Quantum metrology infers the parameter related to distance ranging based on the physical dynamics of the probing system. In this context, we employ quantum bits, known as qubits, to approximate the probes in use. By manually controlling the dynamics of a qubit and utilizing the unique property of quantum entanglement, we expect that the readout from the metrology system would admit a linear combination of the square of distances through the solution of governing Schr\"odinger equation \cite{Lectures_on_QI, QCQI}. While classical ranging methods can algorithmically compute such combinations through repeated measurements, this approach tends to increase overhead and exacerbate measurement errors. In contrast, \rangname exploits the inner tension among entangled qubits, enabling the measurement of sophisticated distance combinations with the same level of error as a single target-anchor distance measurement.

\par A salient feature of \sysname is that by preparing a special probing state corresponding to specific scheme of distance combination, the induced MLE problem would be convexified. Solving the induced optimization would be computationally efficient through computing a least-square (LS) regression \cite{Least_Square}, while the derived solution maintains substantial reliability even when subjected to typical noise conditions. {By a generic configuration of \rangname scheme, $d$ rangings are adequate for \sysname to generate highly dependable position estimation in $d$-dimensional space}. To validate our study, we conducted simulations with noisy distance ranging readouts. The results show that \sysname significantly outperforms baseline approaches using classical ranging, reduces the error metric RMSE by at least 73\% and the average time consumption by at least 97.6\%, and consistently saturates the theoretical lower bound of RMSE.

\par To the best of our knowledge, we are the first to study localization approaches based on quantum ranging. 
Our research takes a pioneering step towards the utilization of quantum-enhanced ranging with quantum entanglement for next-generation localization technologies. We believe this would provide insight to the mobile computing community, and open innovative research opportunities in the fields of both sensor network localization and quantum computing. Our contributions are summarized as follows:

\begin{itemize}
    \item We introduce the problem of range-based quantum localization to the community for the first time and present an analytical formulation of a novel localization approach using quantum-enhanced ranging. 
    \item We propose \rangname, which utilizes the evolutionary dynamics of quantum probe under certain external field manipulation and the methodologies in quantum metrology.
    \item We formulate the localization problem of \sysname under specific probing scheme of \rangname, resulting in a convex optimization problem that can be efficiently solved via weighted least-square regression, and requires only $d$ rangings for localization in $d$-dimensional space. 
    \item We experimentally demonstrate that the proposed \sysname outperforms conventional range-based localization approaches significantly in accuracy and latency, and saturates the Cram\'er-Rao Lower Bound (CRLB) consistently.
\end{itemize}

The rest of the paper is organized as follows. We summarize related works in \S\ref{sec:related_work}, and present a primer on quantum metrology and classical ranging in \S\ref{sec:preliminaries}. We formulate probing particle dynamics in \S\ref{sec:controlled_dynamics}, followed by ranging model in \S\ref{sec:ranging_model} and localization algorithm in \S\ref{sec:localization_by_optimization}. 
\S\ref{sec:numerical_analysis} reports the evaluation results and \S\ref{sec:conclusion} concludes the paper. 


\section{Related Works}
\label{sec:related_work}
\textbf{\textit{Range-based Localization}} Range-based localization has been a subject of intense study, which involves two key problems: ranging and localization. 
Ranging is usually done by reversing the propagation distances from various signals, \eg, GPS, WiFi, mmWave, ultrasound, \etc, with different ranging models, \eg, AoA, TDoA, RSSI, \etc. Many efforts have been made towards localization with certain structure of distance information, including the trilateration-based algorithms, conic relaxation, and MDS-MAP. 


Solving problems induced by trilateration often requires the use of linearization or pseudo-linearization \cite{Chan_Algorithm}, and the performance deteriorates significantly due to inaccurate distance measurements and error accumulation, thus further refinement is required \cite{SDP_localize}. In \cite{Beyond_Trilateration, Noise_Tolerant_Loc}, noise-tolerant trilateration-based algorithms are proposed.
\par Localization by conic relaxation converts non-convex constraints in the problem formulation into convex ones. In \cite{Theory_of_Network_Loc}, So and Ye studied the theory of semidefinite programming (SDP) in sensor network localization, while in \cite{SDP_TDoA_Loc}, Luo \etal applied SDP technique to TDoA localization. Tseng \cite{SOCP_Loc} proposed second-order cone programming (SOCP) method as an efficient variant of SDP. Although relaxation method achieves high accuracy in estimating sensor locations, its complexity is in general not satisfactory \cite{Handbook_of_SDP}, and is thus only applicable to small-scale problems.
\par Multidimensional scaling (MDS) is a special technique aimed at finding low-dimensional representations for high-dimensional data. MDS-MAP \cite{MDS_MAP} constructs a relative map through distance matrix, and localizes, nodes by transforming the map into a absolute map with sufficient and accurate distance measurements.

\change \textbf{\textit{Quantum Metrology}} Quantum metrology \cite{Beat_quantum_limit, quantum_metrology, AdvQuantumMetrology} emerged as an increasingly important research area, where quantum entanglement and coherence are harnessed to boost the precision of sensing beyond the limit of classical sensors in various fundamental scenarios, including thermometry \cite{Quantum_Thermometry}, reference frame alignment \cite{Quantum_Reference_Frame_Transmission}, and distance measurement \cite{Quantum_Positioning}. Controlled evolution of quantum system is also widely studied, largely based on the control theory so as to create certain state evolution in realizing different sensing tasks \cite{Quantum_Control_Bound_State, Effect_Control_Open_Quantum_Sys}. Besides theoretical works, a primitive quantum sensor network has been lately implemented \cite{quantum_sensor_network}, while recent experiment has demonstrated the feasibility of generating large Greenberger–Horne–Zeilinger (GHZ) state \cite{Generate_GHZ_state}. Experimental works demonstrate the feasibility to prepare widely-used probes in quantum metrology, including the ones utilized by \sysname.

\change \textbf{\textit{Quantum-assisted Localization}} There is no significant amount of work presented in the interdisciplinary field of quantum information and localization. A few existing works enhance fingerprint-based localization by accelerating computation in fingerprint database searching using quantum algorithms. Grover in \cite{Quantum_Search} improves the asymptotic time of searching in an unstructured dataset from $O(n)$ to $O(\sqrt{n})$. Buhrman \etal \cite{Quantum_Fingerprinting} introduces the concept of quantum fingerprints and proves its exponential improvement in storage complexity compared to classical one. Subsequent works include the quantum fingerprint localization \cite{Quantum_Algorithm_RF_Localization}, two-stage transmitter localization method with quantum sensor network \cite{Quantum_Sensor_Network_Localization}, and machine learning-based WiFi sensing localization augmented with quantum transfer learning \cite{Quantum_Transfer_Learning_Wifi_Sensing}. 
To our awareness, there is no prior work on range-based localization with quantum ranging.


\section{Preliminaries}
\label{sec:preliminaries}

\subsection{Range-based Localization Model}
\label{sec:ranging_loc}
A typical range-based localization model on a $d$-dimensional space where $d \in \{2, 3\}$ consists of $n$ nodes with accurate positions $\texttt{Anc} = \{\bm{a}_1, \dots, \bm{a}_n \}$ fixed on $\mathbb{R}^d$ under arbitrary topology, called the \textit{anchors}. We consider an \textit{idealized} picture of localization, where all facilities involved have full knowledge of the correspondence and localization of all available anchors. A sensor at position $\bm{x} \in \mathbb{R}^d$ communicates with a subset of $\texttt{Anc}$ through certain medium and acquires information on the functionals of sensor-anchor distances, denoted as $\mathcal{S}(\{ d_i:~ i \in I \subset \{1, \dots, n\} \} | \bm{\vartheta})$ where $I$ is an index set indicating the indices of utilized anchors, $d_i := \|\bm{x} - \bm{a}_i \|$ where $\|\cdot\|$ is the Euclidean norm on $\mathbb{R}^d$, and $\bm{\vartheta}$ parameterizes the ranging process. We herein refer to this process as \textit{ranging}. Moreover, we assume the anchors and sensor positions are bounded, \ie, there exist scalars $0 < \kappa_a, \kappa_s < \infty$, such that for $1 \leq i \leq n$, $\|\bm{a}_i\|_{\infty} \leq \kappa_a$ and $\|\bm{x}\|_{\infty} \leq \kappa_s$ where $\|\cdot\|_{\infty}$ stands for the vector infinite norm with $\|\bm{v}\|_{\infty} := \max_i |v_i|$. Suppose a total of $m$ rangings are available, the objective of the sensor is to fully exploit available information $\{\mathcal{S}^{(k)} \}_{k=1}^m$ where the superscript $k$ specifies the signal acquired from the $k^{\text{th}}$ ranging, and estimate its position $\hat{\bm{x}} \in \mathbb{R}^d$.

\subsection{Comparing Classic Ranging and \rangname}

\par Conventional range-based localization protocols have different distance measurement scenarios, each corresponds to a specific form of signal-distance mapping $\mathcal{S}(\cdot)$. The majority of such mapping involves one or two anchors, including the angle of arrival (AoA), 
$
    \mathcal{S}_{\text{AoA}}(d_i | \theta, \lambda) = 2\pi \frac{\cos \theta}{\lambda} d_i
$
where $\mathcal{S}_{\text{AoA}}$ is the phase difference of adjacent antennas; the time of arrival (ToA), $\mathcal{S}_{\text{ToA}}(d_i|v) = \frac{2d_i}{v}$ where $\mathcal{S}_{\text{ToA}}$ is the time difference between signal emission and recapture; the time differences of arrivals (TDoA),
$
\mathcal{S}_{\text{TDoA}}(d_i, d_j | v) = \frac{1}{v} \left|d_i - d_j\right|
$
where $\mathcal{S}_{\text{TDoA}}$ is the time differences of arrival at the paired and synchronized sensors; and the received signal strength indicator (RSSI),
$
\mathcal{S}_{\text{RSSI}}(d_i | P_t, G_t, G_r, \lambda) = \frac{P_t G_t G_r \lambda^2}{16\pi^2 d_i^2}
$
where $\mathcal{S}_{\text{RSSI}}$ is the received signal power at the sensor \cite{Sensor_Loc}.
\par A notable drawback of classic ranging is the requirement for target-anchor distances to be measured sequentially and individually, as in the cases of AoA, ToA, and RSSI, or in pairs for TDoA. Large numbers of ranging would be imperative if full utilization of anchors is required. This sequential measurement process not only introduces extra complexity and overhead of ranging into the localization task but also results in the system's vulnerability to noise and environmental fluctuations, as the overall effect of normal noise integrated in distance $d_i$ to the eventual solution would be unpredictable. Moreover, substituting the primitive form of $\mathcal{S}(\cdot)$ into the localization problem arising from either MLE or error minimization \cite{SDP_localize} always introduces computationally expensive optimization problems \cite{Noise_Tolerant_Loc}. 

In contrast, the \textit{Quantum-Enhanced Ranging (\rangname)} emerges as an innovative advancement, endowed with a unique capability in settling both issues. It enables the simultaneous ranging of a special combination of distances between a target sensor and an arbitrary number of anchors within a single physical measurement, which is nearly impossible to achieve in classical systems and allows a convexified localization problem. In the following section, we will present a primer on quantum metrology, which underpins \rangname, while leaving the analysis of the exact form of the proposed quantum-enhanced ranging $\mathcal{S}_{\sysname}$ to \S\ref{sec:QuER}. 



\subsection{Quantum Metrology}
\label{sec:quantum_metrology}

The proposed \rangname is based on quantum metrology. We first briefly introduce the principles of quantum metrology and present its generic readout scenarios. 

\subsubsection{Quantum Metrology for Parameter Measurement}
Quantum metrology targets measuring physical parameters with high precision with the aid of quantum mechanics principles \cite{quantum_metrology}. It typically includes \textbf{\textit{(i)}} Preparing a \textit{probe}, described by a quantum state $\ket{\psi}$ in the environment with underlying \textit{Hilbert space} $\mathcal{H}$, which under an orthornormal basis $\{\ket{n} \}$ of $\mathcal{H}$ can be expressed as $\ket{\psi} = \sum_n a_n \ket{n}$ and physically exhibits state $\ket{n}$ with probability $|a_n|^2$, subject to $\sum_n |a_n|^2 = 1$ \cite{QCQI}; \textbf{\textit{(ii)}} Letting it interact with external system, which can be represented by a \textit{unitary transformation} $\mathcal{U}_\phi$ encoded with targeted parameter set $\phi$; and \textbf{\textit{(iii)}} Extracting information on $\phi$ by quantum measurement, specified by a set of \textit{measurement operators} $\{\Pi_i \}_{i \in \mathbb{N}}$. 
\par The probe state and measurement operators may assume to be either separable, or \textit{entangled} \cite{quantum_metrology}, corresponding to whether it is feasible to find the decomposition $\ket{\psi} = \ket{\psi_1} \otimes \cdots \otimes \ket{\psi_n}$, where $\ket{\psi_i} \in \mathcal{H}_i$ is the state in subspace $\mathcal{H}_i \subset \mathcal{H}$, and $\otimes$ represents the tensor product operation. \sysname employs this procedure as a subroutine to decode multiple sensor-anchor distances information from the entangled state with one-shot ranging.

\subsubsection{Generic Readout Scenario of Quantum Metrology}
\label{sec:readout_scenario}
A generic framework of an atomic probing system is encompassed by the following: Practically, the probe is prepared as a uniform superposition $\ket{\psi_{\text{init}}} = \frac{1}{\sqrt{2}}\left(\ket{a} + \ket{b}\right)$, where $\ket{a}, \ket{b}$ are arbitrary orthonormal states in the space $\mathcal{H}$ \cite{quantum_metrology, Quantum_Theory_Phase_Estimation}. Applying the parameterized unitary operation $\mathcal{U}_{\phi}$ on $\ket{\psi_{\text{init}}}$ yields the phase state \cite{AdvQuantumMetrology}:
\begin{equation}
    \ket{\psi_{\phi}} = \frac{1}{\sqrt{2}}\left(\mathcal{U}_\phi \ket{a} + \mathcal{U}_\phi \ket{b}\right) \propto \frac{1}{\sqrt{2}} \left( \ket{a} + e^{-i\phi} \ket{b} \right).
\end{equation}
\par Let $\Pi := \ket{\psi_{\text{init}}} \bra{\psi_{\text{init}}}$ denote the \textit{projection operator} on subspace spanned by the probe, we apply the positive operator-valued measurement (POVM) \cite{QCQI} on $\ket{\psi_{\phi}}$, specified by a couple $\{\Pi, \bm{1}_{\mathcal{H}} - \Pi \}$ where $\bm{1}_{\mathcal{H}}$ is the identity map on $\mathcal{H}$. The readout process involves verifying whether $\ket{\psi_\phi}$ resides in the subspace of $\ket{\psi_{\text{init}}}$. The outcome would simply be either `yes' (encoded as 0) or `no' (encoded as 1), associated with probabilities
\begin{equation}
    \label{eqn:probability_outcome} 
    \left\{\begin{aligned}
    &\text{Pr}(\text{outcome} = 0) = \text{Tr} \left( \Pi \ket{\psi_{\phi}} \bra{\psi_{\phi}}  \right) = \cos^2 \frac{\phi}{2}, \\
    &\text{Pr}(\text{outcome} = 1) = 1 - \text{Pr}(\text{outcome} = 0) = \sin^2 \frac{\phi}{2}. \end{aligned} \right.
\end{equation}
Repeating the procedures allows us to analyze the value of $\phi$ with statistical tools such as maximum likelihood estimator and Bayesian inference \cite{Quantum_Theory_Phase_Estimation}.
\par In particular, when $\ket{a}, \ket{b}$ are in $N$-tensor form, \ie, $\ket{a} = \ket{a_1} \otimes \cdots \otimes \ket{a_N}$ and $\ket{b} = \ket{b_1} \otimes \cdots \otimes \ket{b_N}$, while the operator $\mathcal{U}_\phi$ admits decomposition $\mathcal{U}_{\phi} = \mathcal{U}_{\phi_1} \otimes \cdots \otimes \mathcal{U}_{\phi_N}$, as the subscripts index the subsystem $\mathcal{H}_i$ the quantum states and operators live on. By nature of tensor product, relative phase can thus be alternatively expressed as the sum of relative phases in subsystems, specified by:
\begin{equation}
\begin{aligned}
    \mathcal{U}_{\phi} \ket{\psi_{\text{init}}} &= \frac{1}{\sqrt{2}}\left(\mathcal{U}_{\phi_1} \ket{a_1} \otimes \cdots \otimes \mathcal{U}_{\phi_N} \ket{a_N} \right.\\
    &+   \left.\mathcal{U}_{\phi_1} \ket{b_1} \otimes \cdots \otimes \mathcal{U}_{\phi_N} \ket{b_N} \right) \\
    &\propto \frac{1}{\sqrt{2}} \left( \ket{a_1} \otimes \cdots \otimes \ket{a_N} + e^{-i\phi_1} \ket{b_1} \otimes \cdots \otimes e^{-i\phi_N} \ket{b_N} \right) \\
    &= \frac{1}{\sqrt{2}} \left(\ket{a} + e^{-i\sum_{j=1}^N \phi_j } \ket{b} \right).
\end{aligned}
\end{equation}
Quantum metrology utilizes the above phase accumulation phenomenon to improve the asymptotic error by an $N^{-1/2}$ factor when detecting physical quantities, compared to classical metrology system \cite{AdvQuantumMetrology}. \rangname operates on an alternative advantage of this unique entanglement property. By deliberately correlate each relative phase $\phi_j$ with the particle's travel distance, or equivalently, its time-of-flight (ToF), it enables multiple distances information to be encoded into the joint relative phase $\phi := \sum_{j=1}^N \phi_j$. The following sections \S\ref{sec:controlled_dynamics} and \S\ref{sec:ranging_model} will elaborate on the specific time-dependent evolution of a unique quantum state under certain external controls that \rangname would use.

\section{Controlled Dynamics of A Qubit}
\label{sec:controlled_dynamics}
The controlled electrodynamics of quantum particles under the theory of quantum mechanics is crucial to the realization of \rangname and our proposed \sysname. For an isolated physical system with Hamiltonian $H$, the dynamic of any time-dependent quantum state $\ket{\psi(t)}$ in the Hilbert space is governed by the following \textit{Schr\"odinger equation} \cite{Lectures_on_QI},
\begin{equation}
    \label{eqn:Schordinger}
    i \hbar \ppt \ket{\psi(t)} = H \ket{\psi(t)},
\end{equation}
where $\hbar$ is the reduced Planck constant, $i := \sqrt{-1}$, and $\partial / \partial t$ is the partial derivative operator with respect to time. 
Specifically, we consider a two-level approximation \cite{Lectures_on_QI} of an arbitrary particle, where only two energy levels are considered among multiple possible energy states. The two-level system includes a state of the lowest energy level, called the \textit{ground state} with notion $\ket{g}$ and energy $E_g$, and a state with energy increased through energy absorption with external circumstance, called the \textit{excited state} with notion $\ket{e}$ and energy $E_e$. The energy difference can be expressed as $\Delta E = E_e - E_g = \hbar \omega_0$, where $\omega_0$ is the particle frequency according to the theory of Louis de Broglie \cite{Lectures_on_QI, QCQI}.

\begin{figure}[t]
    \centering
    \includegraphics[width=0.8\linewidth]{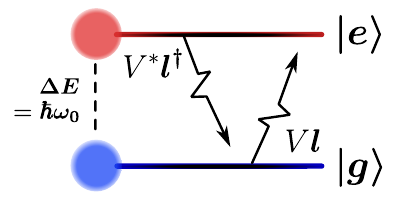}
    \caption{Dynamics of qubit with coupled energy levels}
    \label{fig:energy_jump}
\end{figure}

\par The external electromagnetic field can be viewed as a mechanism coupling the two energy levels, resulting in some implicit transitions between them. Precisely, we denote $\dagger$ to be the conjugate transpose of operators and states, $\bra{\psi} := \ket{\psi}^\dagger$, $\bm{l} := \ket{e}\bra{g}$ the atomic laddering operator \cite{Lectures_on_QI} transiting the ground state to the excited one, and $\bm{l}^\dagger = \ket{g} \bra{e}$ the atomic descending operator acting the opposite, as shown in \fig \ref{fig:energy_jump}. Then the exact manner of such a coupling mechanism can be expressed as
$
    V \bm{l} + \left( V \bm{l} \right)^\dagger
$
\cite{QCQI}, where $V \in \mathbb{C}$ is a complex scalar function characterizing the coupling behaviour. For conciseness, we choose $E_0 = \hbar \omega_0 / 2$ to be the energy zero level, thereby the Hamiltonian of the coupled system \cite{Lectures_on_QI} can be formulated by:
\begin{equation}
\begin{aligned}
    H = \frac{\hbar \omega_0}{2} \left( \bm{l} \bm{l}^\dagger - \bm{l}^\dagger \bm{l}  \right) + V \bm{l} + V^* \bm{l}^\dagger,
\end{aligned}
\end{equation}
where $V^*$ represents the conjugate of complex number $V$.
\par The two-level approximation enables us to encode the states $\{\ket{e}, \ket{g} \}$ into a single qubit by defining $\ket{0} := \ket{e}$ and $\ket{1} := \ket{g}$, and $\{\ket{0},\ket{1} \}$ would form an orthornormal basis of underlying Hilbert space. To investigate how the qubit would evolve when the external mechanism is manually controlled, we hereby consider the time-dependent coupling $V(t) = \epsilon(t) e^{i\theta(t)}$, where $\epsilon, \theta$ are coupling magnitude and field spinning rate respectively, both are real functionals on the time horizon $\mathcal{T} = [0, \infty)$. Analytical intractability of solving the Schr\"ordinger equation with time-dependent Hamiltonian can be settled by a separation of operator: Consider the decomposition of $H$ \cite{Quantum_Perturbation_Theory} as
$
    H = H_0 + D(t)
$, where $H_0$ is a time-independent full-rank operator (\ie, $\text{rank}H_0 = \dim \mathcal{H}$), and $D(t)$ incorporates time-dependent terms. Suppose $H_0$ admits spectrum $\{E_j\}$ with eigenstates $\{\ket{j} \}$, then by assuming $\ket{\psi(t)} = \sum_j c_j(t) e^{-iE_j/\hbar t} \ket{j}$ with $c_j(t)$ being undetermined time-dependent coefficients, constraints on $c_j(t)$ can be derived in light of \eqref{eqn:Schordinger}:  
\begin{equation}
    \label{eqn:coeffcient_time}
    i\hbar  \ppt c_j(t) = \sum_{k} c_k(t) \exp\left\{ -\frac{i}{\hbar} (E_k - E_j) t \right\} \bra{j} D(t) \ket{k}.
\end{equation}
Applying \eqref{eqn:coeffcient_time} to our proposed case, set $H_0 = \frac{\hbar \omega_0}{2} \left(\bm{l}\bm{l}^\dagger - \bm{l}^\dagger \bm{l} \right)$, and $D(t) = V(t) \cdot \bm{l} + V^*(t) \cdot \bm{l}^\dagger$, the coefficients $c_0(t), c_1(t)$ should satisfy
\begin{equation}
\label{eqn:coupled_eqn}
    \left\{ \begin{aligned}
            &\ppt c_0(t) = -\frac{i}{\hbar} \epsilon(t) e^{i(\theta(t) + \omega_0 t)} c_1(t), \\
            &\ppt c_1(t) = -\frac{i}{\hbar} \epsilon(t) e^{-i(\theta(t) + \omega_0 t)} c_0(t).
    \end{aligned} \right.
\end{equation}
Formulating the above coupled differential equations would yield the following equation:
\begin{equation}
\label{eqn:coefficients}
\begin{aligned}
    \frac{\partial^2}{\partial t^2} c_j(t) &- \left\{\frac{1}{\epsilon(t)}\ppt \epsilon(t) + i \cdot (-1)^j  \left(\ppt \theta(t) 
    + \omega_0 \right)   \right\} \ppt c_j(t) \\
    &+ \frac{\epsilon^2(t)}{\hbar^2} c_j(t) = 0, \quad j \in \{0, 1\}.
\end{aligned}
\end{equation}
A tentative solution would be $c_j(t) \propto e^{i\eta_j(t)}$ where $\eta_j$ is a real funtional on $\mathcal{T}$. Substituting it into \eqref{eqn:coefficients} yields following equation:
\begin{equation}
\begin{aligned}
    \left(\ppt \eta_j (t) \right)^2 &- (-1)^{j} \left( \ppt \theta(t) + \omega_0 \right) \ppt \eta_j(t) - \frac{\epsilon^2(t)}{\hbar^2} \\
    &= i \left( \frac{\partial^2}{\partial t^2} \eta_j(t) - \frac{\ppt \epsilon(t)}{\epsilon(t)} \ppt \eta_j(t)  \right), \quad  j \in \{0, 1\}.
\end{aligned} 
\end{equation}
Since $\eta_j, \epsilon, \phi$ are all real functionals, the coincidence of two sides in above equation demonstrates the following constraints on the form of quantum state:
\begin{equation}
\label{eqn:dynamic_constraint}
  \begin{array}{cl}
\ket{\psi(t)} = &\sum_{j \in \{0, 1\} } \sum_{\eta_j} Q_{\eta_j} e^{i(\eta_j(t) - (-1)^{j} \omega_0 t / 2)} \ket{j}, \\
\text{s.t.}  &\left(\ppt \eta_j (t) \right)^2 - (-1)^{j} \left( \ppt \phi(t) + \omega_0 \right) \ppt \eta_j(t) \\
 &- \frac{\epsilon^2(t)}{\hbar^2} = 0, \\
 &\left(\ppt \eta_j(t)\right)^{-1} \frac{\partial^2}{\partial t^2} \eta_j(t) = \epsilon^{-1}(t) \ppt \epsilon(t),
\end{array}
\end{equation}
where $Q_{\eta_j} \in \mathbb{C}$ are complex coefficients, and the inner summations are taken on all possible functionals $\eta_j$. The exact behaviour of state entries can be solved with full knowledge of its initial state $\ket{\psi(0)}$ and coupling factors $(\epsilon(t), \phi(t))$.

\section{Quantum-Enhanced Ranging}
\label{sec:ranging_model}
In this section, we discuss in detail how our \sysname takes advantage of a special case of the above constrained probe qubit evolution. Within the region of localization, we deploy a meticulously controlled field with $\epsilon(t) = \nu(2 \gamma t + \omega_0)$ and $\theta(t) = \gamma t^2$, where $\nu, \gamma > 0$ are positive parameters. We further assume that $\nu \gg \hbar$.

\subsection{Behaviour of Qubit with Uniform Superposition}
\label{subsec:single_qubit_dynamics}
We begin with illustrating the evolutionary behaviour of a qubit $\ket{\psi(0)}$ with uniform quantum superposition state, \ie, it admits equal probability on both of its energy states. It can be prepared by implementing the Hadamard transformation \cite{QCQI} $\mathscr{H}$ on $\ket{0}$: 
\begin{equation}
    \ket{\psi(0)} = \mathscr{H} \ket{0} = \frac{1}{\sqrt{2}} \left(\ket{0} + \ket{1}\right).
\end{equation}
From the expression of $\epsilon, \theta$, we could arrive at an expression of the time variation of single-qubit state $\ket{\psi(t)}$:
\begin{equation}
    \ket{\psi(t)} = c_0(t) e^{-\omega_0 t / 2} \ket{0} + c_1(t) e^{\omega_0 t / 2} \ket{1},
\end{equation}
subject to the initial state consistency and probability completeness
\begin{equation}
\label{eqn:constraints}
\begin{aligned}
    c_0(0) &= c_1(0) = \frac{1}{\sqrt{2}}, \quad \left|c_0(t)\right|^2 + \left|c_1(t)\right|^2 = 1, \quad j = 0, 1, \\
    c_j(t) &= A_j \exp \left\{ i \cdot \frac{(-1)^j + \sqrt{1 + 4\nu^2 / \hbar^2}}{2} (\gamma t^2 + \omega_0 t)  \right\} \\
    &+ B_j \exp \left\{ i \cdot \frac{(-1)^j - \sqrt{1 + 4\nu^2 / \hbar^2}}{2} (\gamma t^2 + \omega_0 t)  \right\}.
    \end{aligned}
\end{equation}
Denote $\Re(C)$ and $\Im(C)$ as the real and imaginary part of a complex number $C \in \mathbb{C}$. Solving the undetermined coefficients $A_j, B_j \in \mathbb{C}$ for $j = 0, 1$ in \eqref{eqn:constraints} subject to \eqref{eqn:coupled_eqn}, we discover that
\begin{equation}
    \label{eqn:sol_coefficients}
    \left\{
    \begin{aligned}
        &\Re(A_0) = \frac{1}{\sqrt{2}} \frac{1 - \tau}{1 + \tau^2}, \quad \Re(B_0) = \frac{1}{\sqrt{2}} \frac{\tau^2 + \tau}{1 + \tau^2}, \\
        &\Re(A_1) = \frac{1}{\sqrt{2}} \frac{\tau^2 - \tau}{1 + \tau^2}, \quad \Re(B_1) = \frac{1}{\sqrt{2}} \frac{\tau+1}{1 + \tau^2}, \\
        &\Im(A_0) = \Im(A_1) = \Im(B_0) = \Im(B_1) = 0, \\
        &\tau = \frac{2\nu / \hbar}{1 + \sqrt{1 + 4\nu^2 / \hbar^2}}.
    \end{aligned}
    \right.
\end{equation}
By denoting $\Delta(t) := -\sqrt{1 + 4\nu^2 / \hbar^2}\left(\gamma t^2 + \omega_0 t\right)$, the previous assumption $\nu \gg \hbar$ indicates that $2\nu / \hbar \gg 1$, and consequently $\tau \to 1$. 
Note that whenever $|\Re(e^{i\Delta(t)})| = \left|\cos(\Delta(t))\right| \gg \frac{1-\tau}{\tau^2 + \tau} $, we have two approximate relations $A_0 + B_0 e^{i\Delta(t)} \approx \frac{1}{\sqrt{2}} e^{i\Delta(t)}$ and $A_1 + B_1 e^{i\Delta(t)} \approx \frac{1}{\sqrt{2}} e^{i\Delta(t)}$. When $\Delta(t) = \pm \frac{\pi}{2} + \varepsilon$ for some scalar $|\varepsilon| \ll 1$ within a period $2\pi$ of the $\cos(\cdot)$ function, $|\cos(\Delta(t))| = |\sin(\pm \frac{\pi}{2} - \Delta(t))| = |\sin(\varepsilon)| \approx |\varepsilon|$. Thus, when $\varepsilon$ satisfies $\frac{1-\tau}{\tau^2+\tau} \ll |\varepsilon| \ll 1$, \eg, $\varepsilon = \sqrt{\frac{1-\tau}{\tau^2+\tau}}$, the following approximation of probe state dynamic can be applied except for intervals $\left[\pm \frac{\pi}{2} - |\varepsilon|, \pm \frac{\pi}{2} + |\varepsilon|\right]$ within a single period: 
\begin{equation}
    \label{eqn:approx_evolution}
    \begin{aligned}
        \ket{\psi(t)} &= \frac{1}{\sqrt{2}}{e^{-\frac{i \omega_0}{2} t} e^{i \cdot \frac{1 + \sqrt{1 + 4\nu^2 / \hbar^2}}{2} \left(\gamma t^2 + \omega_0 t\right)}}  \left( \frac{1-\tau}{1+\tau^2} + \frac{\tau^2 + \tau}{1+\tau^2} e^{i\Delta(t)} \right) \ket{0} \\
        &+ \frac{1}{\sqrt{2}}{e^{\frac{i\omega_0}{2} t}  e^{i\cdot \frac{-1+\sqrt{1 + 4\nu^2 / \hbar^2}}{2} (\gamma t^2 + \omega_0t)}} \left( \frac{\tau^2 - \tau}{1 + \tau^2} + \frac{\tau+1}{1 + \tau^2} e^{i\Delta(t)}  \right) \ket{1} \\
        &\xrightarrow{\tau \to 1,~ |\cos(\Delta(t))| \gg \frac{1-\tau}{\tau^2 + \tau}} ~ \underbrace{e^{\frac{i}{2} \gamma t^2} e^{\frac{i}{2} \Delta(t)}}_{\text{global phase}} \left( \frac{ \ket{0} + e^{-i\gamma t^2} \ket{1}}{\sqrt{2}}  \right) \\
        &\propto \frac{\ket{0} + e^{-i\gamma t^2} \ket{1}}{\sqrt{2}}.
    \end{aligned}
\end{equation}

\begin{figure}[t]
    \centering
    \includegraphics[width=\linewidth]{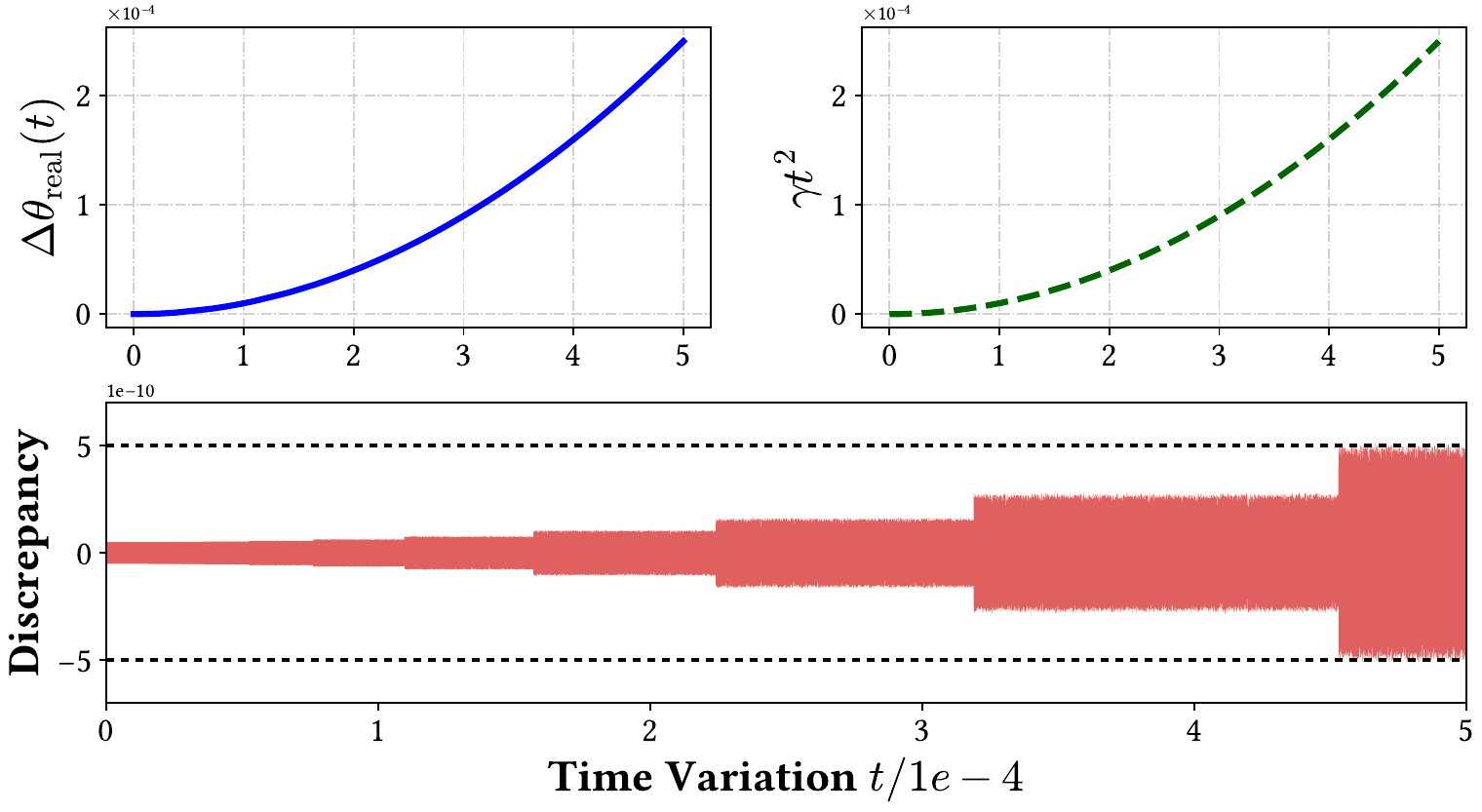}
    \caption{Comparison of real relative phase $\Delta \theta_{\text{real}}(t)$ and $\gamma t^2$. {\rm Minor outliers ($50$ out of $5 \times 10^6$ data points) are filtered out. Here we set $\gamma =10^3 ~\text{rad}/\text{sec}^2$, $\omega_0 = 10^{-2}~\text{rad}/\text{sec}$, and $\nu / \hbar = 10^{10}$. $\frac{1-\tau}{\tau^2 + \tau} = 2.5 \times 10^{-11}$. Absolute discrepancy is bounded by $5 \times 10^{-10}$ while relative phase is on the order of $10^{-4}$. Parameters can be adjusted subject to prior estimation on the order of probe ToF.}}
    \label{fig:simu_theo_compare}
\end{figure}

Such approximation is feasible with a high probability of $1 - \frac{2|\varepsilon|}{\pi}$, and its validity over a certain time period is demonstrated numerically in \fig \ref{fig:simu_theo_compare}. While the global phase factor could not be statistically observed \cite{QCQI}, the two energy states yield a time-dependent relative phase shift with angular speed proportional to the square of time $t$, which enables the statistical detection of qubit ToF in quadratic form \cite{AdvQuantumMetrology}, as outlined in \S\ref{sec:readout_scenario}.


\subsection{Ranging Model of \sysname}
\label{sec:QuER}

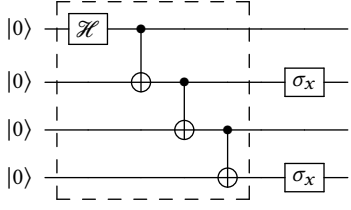
\begin{figure}[t]

$$
\begin{array}{c}
     \Qcircuit @C=1em @R=1em {
     &\lstick{\ket{0}} &\gate{\mathscr{H}} &\ctrl{1} & \qw & \qw &\qw &\qw &\qw \\
     &\lstick{\ket{0}} &\qw &\targ &\ctrl{1} &\qw &\qw &\gate{\sigma_x} &\qw \\
     &\lstick{\ket{0}} &\qw &\qw &\targ &\ctrl{1} &\qw &\qw &\qw  \\
     &\lstick{\ket{0}} &\qw &\qw &\qw &\targ &\qw &\gate{\sigma_x} &\qw     
     \gategroup{1}{3}{4}{6}{1em}{--}
     }
\end{array}
$$

\caption{Quantum circuit of ${\mathscr{E}( \{1, -1, 1, -1\} )}$ applied to $\ket{0}^{\otimes 4}$. {\rm The section outlined by dash line prepares a $4$-qubit GHZ state \cite{QCQI}, while the following $\sigma_x$ gates conduct bit-flipping.}}
\label{fig:quantum_circuit}
\end{figure}

Previous analysis in \S\ref{subsec:single_qubit_dynamics} could be naturally extended to the picture of multi-qubit evolution. This is of great essence to the realization of \sysname. Prior to that, we first outline the settings of our proposed quantum-enhanced ranging. Analogous to classical range-based localization, a \sysname scheme conducts a total of $m$ rangings (\rangname{s}), whereas for the $k^{\text{th}}$ ranging where $1 \leq k \leq m$, the anchors involved would be flexibly identified by an index set $I_k \subset \{1, \dots, n\}$, by recalling that each index $i \in \{1, \dots, n\}$ stands for the $i^{\text{th}}$ anchor available. Each involved anchor $\bm{a}_i$ subject to $i \in I_k$, is assigned a binary-valued parameter $w_{i, k} \in \{-1, 1\}$. $\{w_{i,k}\}_{i \in I_k}$ specifies the \textit{probe scheme} of the $k^{\text{th}}$ \rangname. Accordingly, denote $|\cdot|$ as the set cardinality and $\one{(\cdot)}$ as the indicator function, each ranging would require the following maximally entangled $|I_k|$-qubit probe:
\begin{equation}
\begin{aligned}
    \ket{\psi(0)}_k &= \frac{1}{\sqrt{2}} \sum_{j=0, 1} \bigotimes_{i \in I_k} \ket{\pi_{i, k}^{(j)}}, \\
    \pi_{i, k}^{(j)} &= \one{\{w_{i, k} = -1\}} + j~\text{mod}~2,
\end{aligned}
\end{equation}
which can be prepared by an \textit{entangling operator} $\mathscr{E}_k\left(\{w_{i, k}\}_{i \in I_k}   \right)$ on the input ground state, composed by a sequence of Hadamard gates, controlled-not (CNOT) gates and NOT (Pauli-X) gates \cite{QCQI}. An illustrative example on preparing a four-qubit probe $\frac{1}{\sqrt{2}}\left(\ket{0101} + \ket{1010}\right)$ that corresponds to \rangname scheme $\{1, -1, 1, -1\}$ is shown in \fig \ref{fig:quantum_circuit}.

\par To convexify the localization problem, \sysname further restricts that an even number of anchors are utilized for each ranging process of \rangname (\ie, $|I_k| \in 2 \mathbb{Z}$), and moreover
\begin{equation}
\label{eqn:querloc_requirement}
    \forall k, \quad \sum_{i \in I_k} \one{\{w_{i, k} = 1\}} = \sum_{i \in I_k} \one{\{w_{i, k} = -1\}}.
\end{equation}
\par The sensor triggers each ranging procedure by simultaneously emitting the entangled probe qubits, which subsequently evolve continuously in the controlled external field until getting received by the sensor after being reflected by the specific anchor. Denote $\bm{t}_k = \{t_i \}_{i \in I_k}$ the time instants at which each qubit is retrieved by the sensor, the probe would end up with the following form:
\begin{equation}
    \label{eqn:eventual_state}
    \begin{aligned}
    \ket{\psi(\bm{t}_k)}_k &= \frac{1}{\sqrt{2}} e^{\frac{i}{2} \sum_{i \in I_k} \Delta(t_i) } e^{i \xi_k} \left(  \bigotimes_{i \in I_k} \ket{\pi_{i, k}^{(0)}} + e^{-i \chi_k}\bigotimes_{i \in I_k} \ket{\pi_{i, k}^{(1)}} \right) \\
    &\propto \frac{1}{\sqrt{2}} \left(  \bigotimes_{i \in I_k} \ket{\pi_{i, k}^{(0)}} + e^{-i \chi_k}\bigotimes_{i \in I_k} \ket{\pi_{i, k}^{(1)}} \right),
    \end{aligned}
\end{equation}
where 
\begin{equation}
    \label{eqn:phases}
    \begin{aligned}
        \chi_k = \gamma \sum_{i \in I_k} w_{i, k} t_i^2, \quad \xi_k = \frac{1}{2} \chi_k.
    \end{aligned}
\end{equation}
\par Special quantum properties such as the Zeno effect \cite{Zeno_effect} enable us to inhibit the successive spontaneous evolution once the qubit is returned. Thus, no time synchronization is required among different probing qubits, which is of great concern in traditional ToA and TDoA ranging models \cite{Sensor_Loc}. With the relative phase $\chi$ acquired, by assuming photons are employed as the probes \cite{quantum_metrology}, whose propagation speed is the speed of light $c$, we could use the instantaneous relation between distance and ToF $t_i = 2d_i / c$ to derive the signal-distance mapping $\mathcal{S}_{\sysname}^{(k)}$ for all $k \in \{1, \dots, m\}$,
\begin{equation}
\label{eqn:signal_form}
   \mathcal{S}_{\sysname}^{(k)} \left( \{d_i\}_{i \in I_k} | \{w_{i, k}\}_{i \in I_k}, \gamma, c \right) := \chi_k = \frac{4 \gamma}{c^2} \sum_{i \in I_k} w_{i, k} d_i^2.
\end{equation}
Above non-linear ToF effect is an instance of quantum control that realizes nonlinear quantum dynamics with external field manipulation \cite{Non_Linear_Quantum_State_Evolution, Robust_Quantum_Control}, which is of increasing interest in the field of quantum information processing.
In the next section, we will reformulate the localization task as a simple optimization problem based on above structure of phase information.

\section{Localization via Quantum Ranging}
\label{sec:localization_by_optimization}
\subsection{Reformulating the Phase-Distance Relations}
Upon obtaining the accumulated phase $\chi_k$, we can further expand the terms in previous equality \eqref{eqn:signal_form} as follows:
\begin{equation}
\label{eqn:reformulate}
    \begin{aligned}
        \frac{4 \gamma}{c^2} \sum_{i \in I_k} w_{i, k} d_i^2 &= \frac{4 \gamma}{c^2} \sum_{i \in I_k} w_{i, k} \left(\bm{x} - \bm{a}_i\right)^T \left(\bm{x} - \bm{a}_i\right) \\
        &= \frac{4 \gamma}{c^2} \underbrace{\sum_{i \in I_k} w_{i, k}}_{=~0} \bm{x}^T \bm{x} - \bigg( \frac{8 \gamma}{c^2} \sum_{i \in I_k} w_{i, k} \bm{a}_i \bigg)^T \bm{x} \\
        &+\frac{4 \gamma}{c^2} \sum_{i \in I_k} w_{i, k} \bm{a}_i^T \bm{a}_i = \chi_k, \quad \forall 1 \leq k \leq m.
    \end{aligned}
\end{equation}
The coefficient of $\bm{x}^T \bm{x}$ is eliminated due to the requirement in \eqref{eqn:querloc_requirement}. For mathematical brevity, we apply the following variable substitution after simplifying the equation in \eqref{eqn:reformulate}:
\begin{equation}
    \label{eqn:var_sub}
    \begin{aligned}
    &\bm{L} = \begin{pmatrix}
        \bm{u}_1, \dots, \bm{u}_m
    \end{pmatrix}^T \in \mathbb{R}^{m \times d}, \quad \bm{h} = \begin{pmatrix}
        h_1, \dots, h_m
    \end{pmatrix}^T \in \mathbb{R}^m, \\
    &\bm{u}_k = 2 \sum_{i \in I_k} w_{i, k} \bm{a}_i, \quad h_k =   \sum_{i \in I_k} w_{i, k} \bm{a}_i^T \bm{a}_i - \left(4\gamma\right)^{-1} c^2 \chi_k.
    \end{aligned}
\end{equation}
Finally, by aggregating results from all $m$ distance ranging, the simplification moves us from dealing with a complicated quadratic problem to working with the following elegant and straightforward system of linear equations, which \sysname solves to realize sensor positioning:
\begin{equation}
\label{eqn:prob_statement}
    \mathop{\text{Find}} \quad \bm{x} \in \mathbb{R}^d,  \quad \text{s.t.} \quad \bm{L} \bm{x} = \bm{h}.
\end{equation}

\subsection{Weighted Least-Square Solution}
The systematic bias introduced by the relative phase readout can be modeled as Gaussian noise. It is routine to assume the noise in parallel experiments are independent, yield zero mean, and have standard deviation proportional to the magnitude of observable physical quantities \cite{SDP_localize}. Without loss of generality, we consider all noise are integrated in the scalarized value of signal $\left(4\gamma\right)^{-1} c^2 \chi_k $, which we denote as $\lambda_k$. The noisy measurement readout can be analytically modeled as 
\begin{equation}
\label{eqn:noise_form}
    \widetilde{\lambda}_k = \lambda_k \left( 1 + \delta_k \right), \quad \bm{\delta} \sim \mathcal{N}\left(\bm{0}, \rho^2 \bm{I} \right),
\end{equation}
where $\rho \in [0, 1)$ is a scaling factor that characterizes the extent of measurement error, and $\bm{\delta}$ is a vector of Gaussian noise.
\par We use the $\mathbb{R}^m$ vectors $\bm{\lambda}, \widetilde{\bm{\lambda}}$ respectively to aggregate the exact and noisy measurement readouts. Based on previous assumptions, on observing $\widetilde{\bm{\lambda}}$, the problem \eqref{eqn:prob_statement} can be addressed by solving the following log-likelihood maximization:
\begin{equation}
    \label{eqn:likelihood}
    \begin{aligned}
    \hat{\bm{x}} &=
    \mathop{\arg\max}_{\bm{x} \in \mathbb{R}^d} \quad \log \mathcal{L}\left[ \widetilde{\bm{\lambda}}; \{ w_{i,k} \}_{i \in I_k}:~1 \leq k \leq m;~ \texttt{Anc} \right] \\
    &= \mathop{\arg\max}_{\bm{x} \in \mathbb{R}^d} \quad \log \mathop{\text{Pr}}\limits_{\bm{\delta} \sim \mathcal{N}(0, \rho^2 \bm{I})} \left[ \widetilde{\bm{\lambda}} ~ \bigg|~ \widetilde{\lambda}_k = \lambda_k \left( 1 + \delta_k \right) \right] \\
    &= \mathop{\arg\max}_{\bm{x} \in \mathbb{R}^d} \quad  \sum_{k=1}^m   \log \bigg[\frac{1}{\sqrt{2\pi} \rho \lambda_k } \cdot e^{ \frac{(u_k^T \bm{x} - h_k - \lambda_k + \widetilde{\lambda}_k)^2 }{2\rho^2 \lambda_k^2}} \bigg] \\
    &\approx \mathop{\arg\min}_{\bm{x} \in \mathbb{R}^d} \quad \left\|\sqrt{\widetilde{W}}\left(\bm{L} \bm{x} - \widetilde{\bm{h}}\right)\right\|^2.
    \end{aligned}
\end{equation}
Alternatively, we use term $\widetilde{h}_k = \sum_{i \in I_k} w_{i, k} \bm{a}_i^T \bm{a}_i - \widetilde{\lambda}_k$ as the $k^{\text{th}}$ entry of vector $\widetilde{\bm{h}}$ with noise, and $\widetilde{W} = \text{diag}\left( \widetilde{\lambda}_1^{-2}, \dots, \widetilde{\lambda}_m^{-2}  \right)$ as the diagonal weighting matrix. Approximation in the last equality arises from our insufficient knowledge of the true measurement outcomes $\{\lambda_k\}_{k=1}^m$, and we thus replace them by the noisy observations.
\par Above optimization objective is a typical weighted least square (WLS) problem, which is convex and would yield a closed-form solution \cite{Least_Square}:
\begin{equation}
    \hat{\bm{x}}_{\text{opt}} = \left(\bm{L}^T \widetilde{W} \bm{L}\right)^{-1} \bm{L}^T \widetilde{W} \widetilde{\bm{h}}.
\end{equation}
It is worth noting that solving the \sysname problem requires relatively low computational complexity, as will be discussed in \S\ref{sec:complexity}. Unlike traditional localization methods, our \sysname directly admits a convex optimization problem in its simplified expression and no further transformation is required.

\section{Numerical Analysis}
\label{sec:numerical_analysis}
We present numerical analysis results in this section to demonstrate the performance of \sysname under different testbed settings.
\subsection{Simulation Setups}
\subsubsection{Default Settings of Parameters}
In subsequent experiments, we set up \textit{default values} for a fraction of the parameters included, as listed in \tab \ref{tab:basic_params}.

\begin{table}[htbp]
\caption{Default settings of experiment parameters}
\label{tab:basic_params}
\begin{tabular}{ll}
\hline
\large{\textbf{Parameters}}                     & \large{\textbf{Value}} \\ \hline
Dimension ${d}$            & $3$                             \\
${\kappa_s}$                                         & $100$ (m)                        \\
${\kappa_a / \kappa_s}$                              & $0.5$                           \\
Number of Anchors ${n}$    & $10$                            \\
\texttt{Anc} & $\left\{ \begin{aligned}
    &(0, 0, 0), (\kappa_a, 0, 0), (0, \kappa_a, 0) \\
    &(0, 0, \kappa_a) (\kappa_a, \kappa_a, \kappa_a), (\kappa_a, 0, \kappa_a) \\
    &(\kappa_a, \kappa_a, 0), (0, \kappa_a, \kappa_a), \left(\frac{\kappa_a}{2}, \frac{\kappa_a}{2}, 0\right) \\
    &\left(\frac{\kappa_a}{2}, \frac{\kappa_a}{2}, \kappa_a\right)
\end{aligned} \right\}$           \\
Number of Ranging ${m}$    & $3, 4, 5$                       \\
$\forall k, \quad {|I_k|}$                                            & $2$                             \\
Noise factor ${\rho}$      & $0\% - 5\%$, step length $0.5\%$   \\ \hline
\end{tabular}
\end{table}
In addressing an instance of localization problem , we can, without loss of generality, scale down all distance values by a coefficient, even if these values are of a very large magnitude. Thus, the choice of $\kappa_s$ would not affect the result of our numerical evaluation, and we simply set $\kappa_s = 100$ (m) here. Note that we control the ratio $\kappa_a / \kappa_s$ to be smaller than $1$, so as to generate both near-field and far-field instances, while the latter case is notably more sensitive to ranging noise. Anchors are deployed at the very beginning of the experiment with the specified topology to avoid degeneration of the baseline performance, and remain stationary throughout the simulations. To ensure the feasibility of experimental realization in subsequent works, we employ a $2$-qubit entangled probe in the simulation, and thus $|I_k| = 2$. Entries of signing scheme $\{w_{i, k} \}_{i \in I_k}$ will be specified in the later context, subject to various choices of number of rangings $m$.

\subsubsection{Baseline Algorithms}
We compare \sysname with three range-based localization approaches: \textbf{\textit{(i)}} \textbf{Multilateration + GD}: multilateration is an enhanced version of trilateration to make the solution more robust to noise \cite{Multilateration_Review, Noise_Tolerant_Trilateration} by involving more shots of ranging. We further apply a gradient-descent (GD) refinement \cite{SDP_localize} to the solution of the linear system introduced by multilateration in the presence of noise to provide a convincing baseline. 
\textbf{\textit{(ii)} SDP-based Localization}: introduced to the field of sensor network localization in \cite{Theory_of_Network_Loc}. It is a powerful approach to achieve robust positioning in network with complex topology, we reduce it to the case of single sensor localization.
\textbf{\textit{(iii)}} \textbf{TDoA}: set up the same reference anchor among all time-difference rangings, and locate the sensors by finding the intersection point of a set of elliptic curves with a shared focus. In the following experiment, we use Chan's algorithm \cite{Chan_Algorithm} to settle the non-convexity of distance terms by formulating a pseudo-linear system and solving it with SOCP \cite{Handbook_of_SDP}.

\begin{figure*}[t]
  \begin{minipage}{0.315\linewidth}
    \centering
    \subfloat[RMSE, $m = 3$]{
    \includegraphics[width=\textwidth]{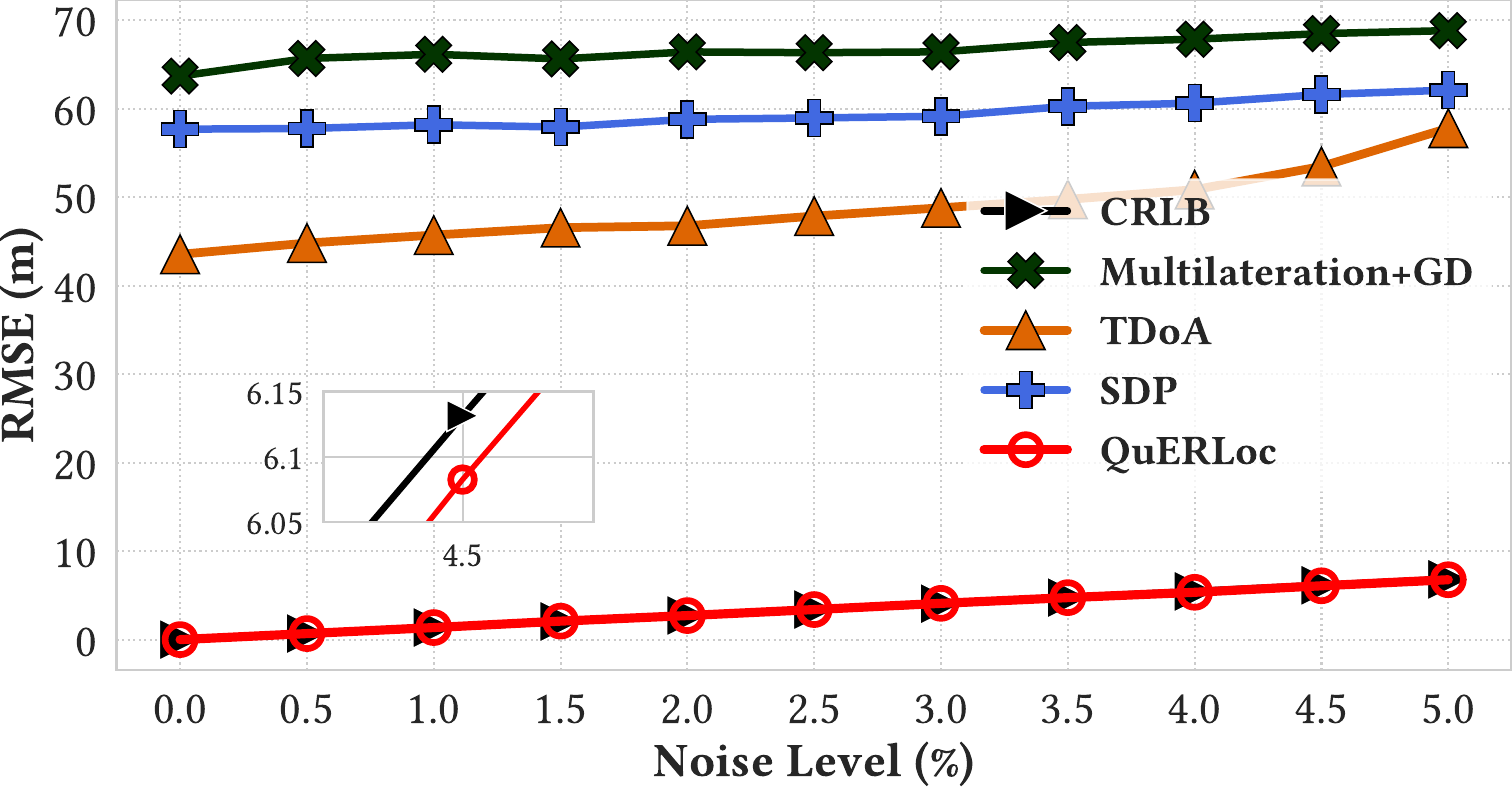}
    \label{fig:rmse_3}
    }
  \end{minipage}
  \hfill
  \begin{minipage}{0.315\linewidth}
    \centering
    \subfloat[RMSE, $m=4$]{
        \includegraphics[width=1.0\textwidth]{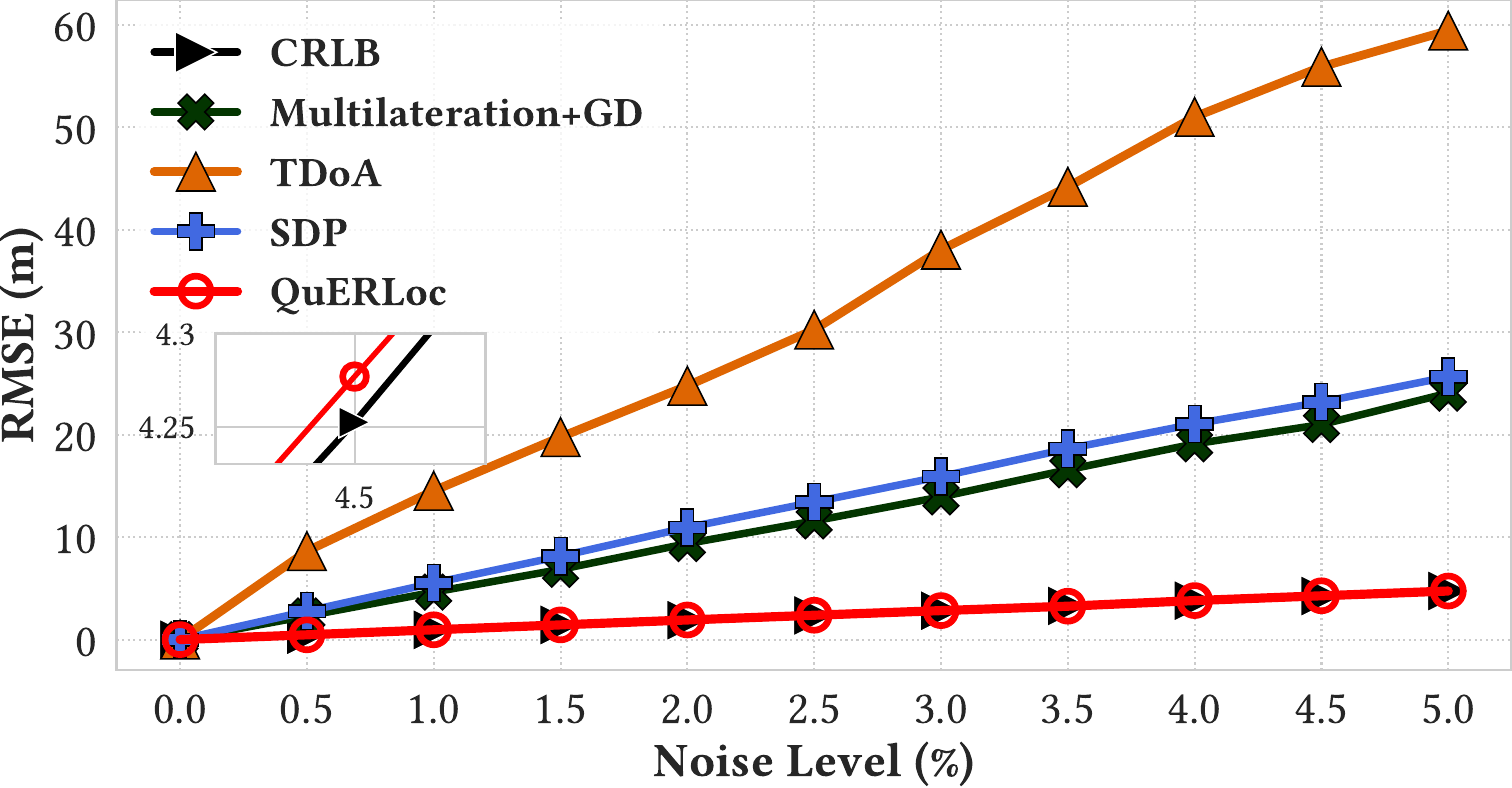}
        \label{fig:rmse_4}
    }
  \end{minipage}  
  \hfill 
  \begin{minipage}{0.315\linewidth}
    \centering
    \subfloat[RMSE, $m=5$]{
    \includegraphics[width=\textwidth]{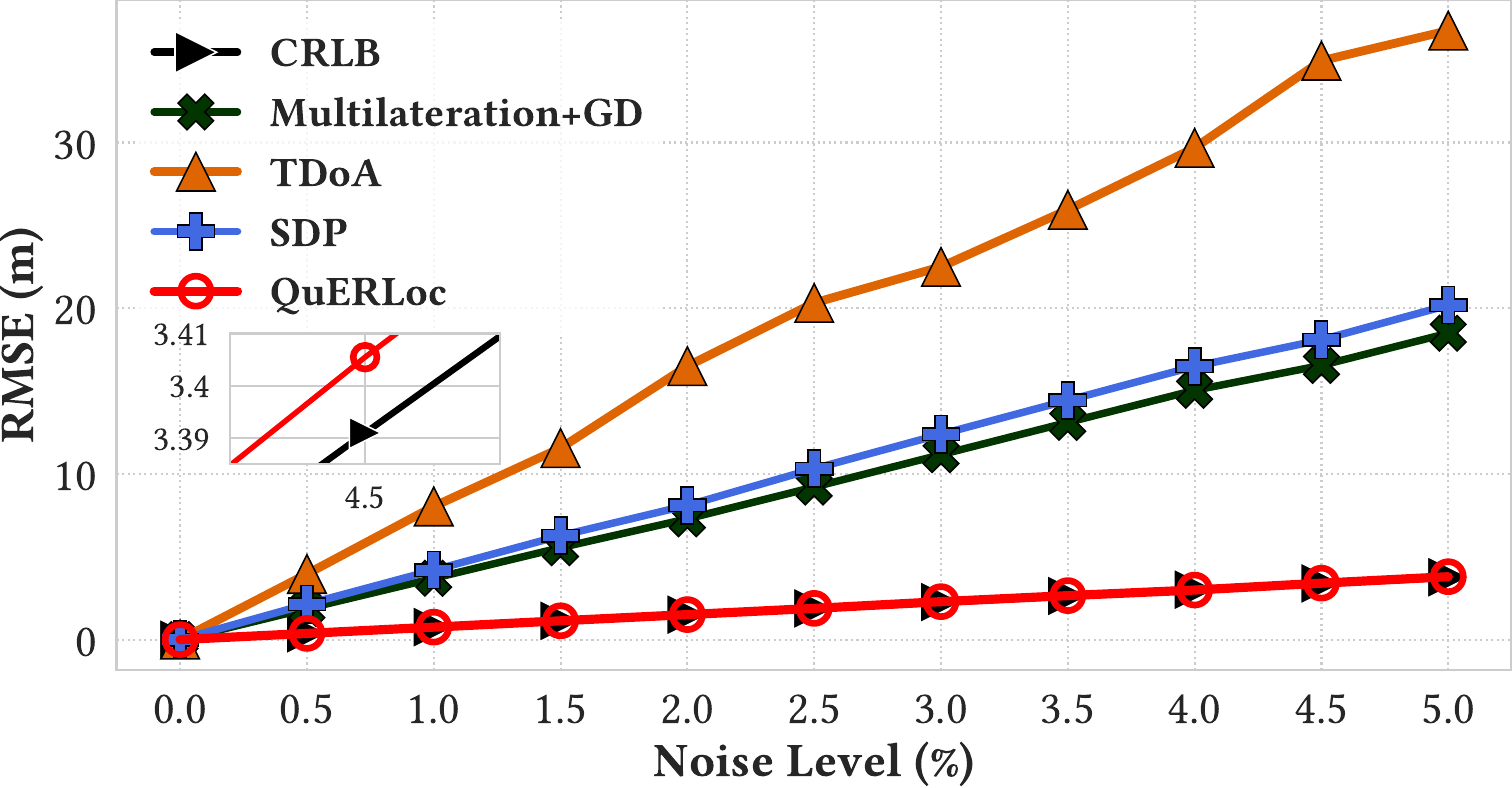}
    \label{fig:rmse_5}
    }
  \end{minipage}
  \begin{minipage}{0.315\linewidth}
      \centering
      \subfloat[CDF, $m=3$]{
      \includegraphics[height= 0.5\textwidth]{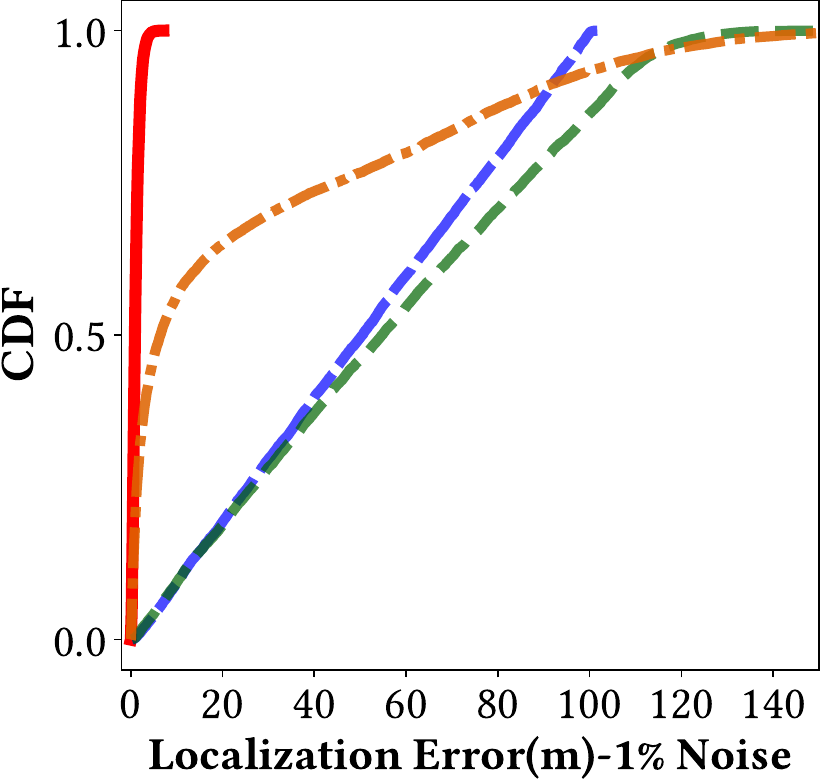}
      \hfill
      \includegraphics[height= 0.5\textwidth]{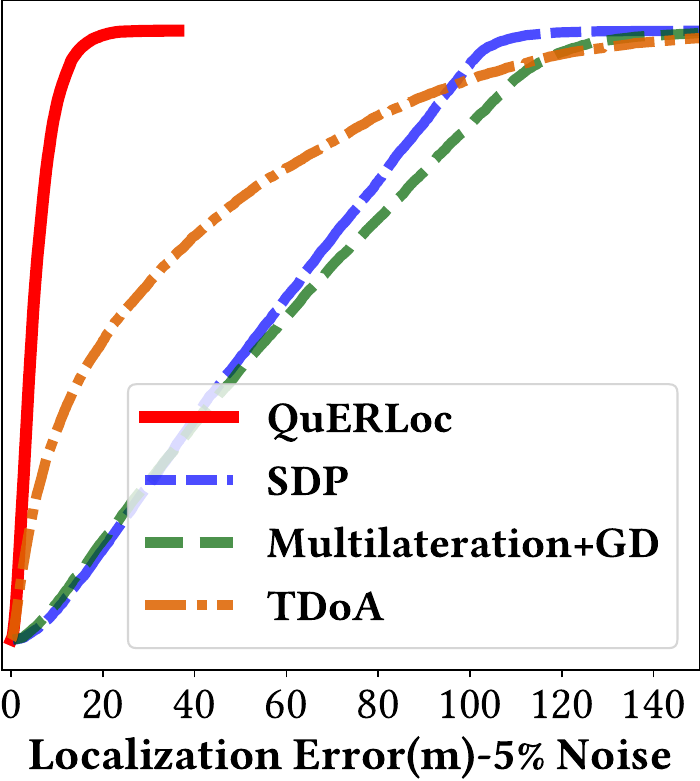}
      \label{fig:cdf_3}
      }
  \end{minipage}
  \hfill 
  \begin{minipage}{0.315\linewidth}
      \centering
      \subfloat[CDF, $m=4$]{
      \includegraphics[height= 0.5\textwidth]{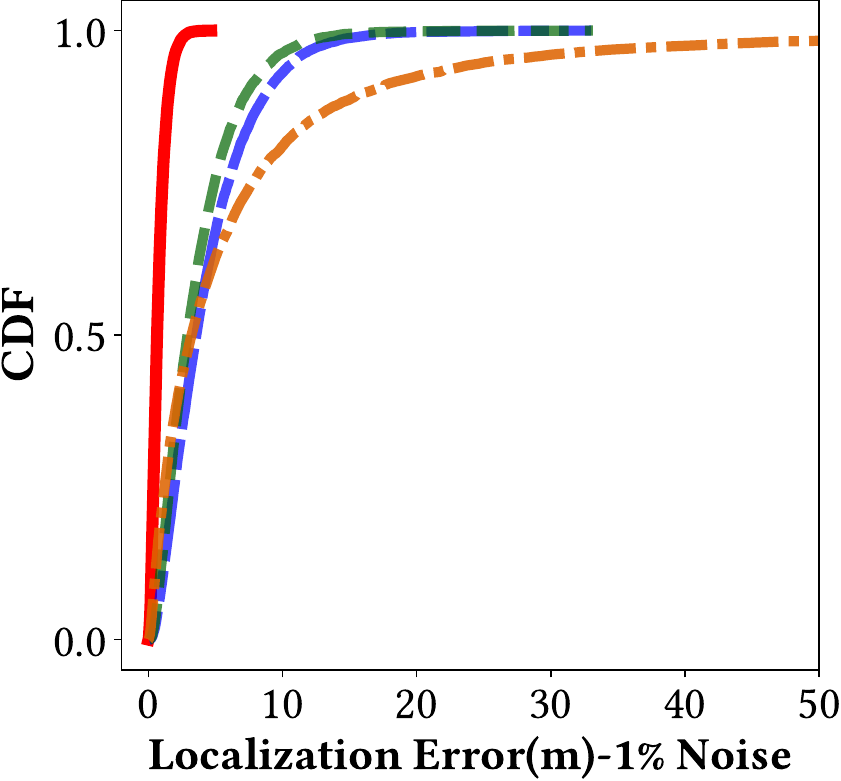}
      \hfill
      \includegraphics[height= 0.5\textwidth]{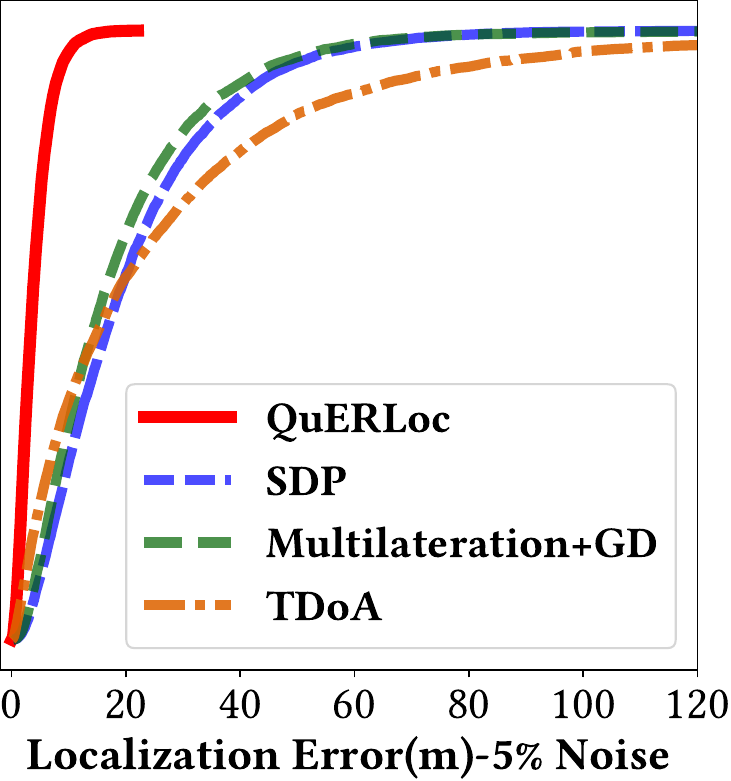}
      \label{fig:cdf_4}
      }
  \end{minipage}
  \hfill 
  \begin{minipage}{0.315\linewidth}
      \centering
      \subfloat[CDF, $m=5$]{
      \includegraphics[height= 0.5\textwidth]{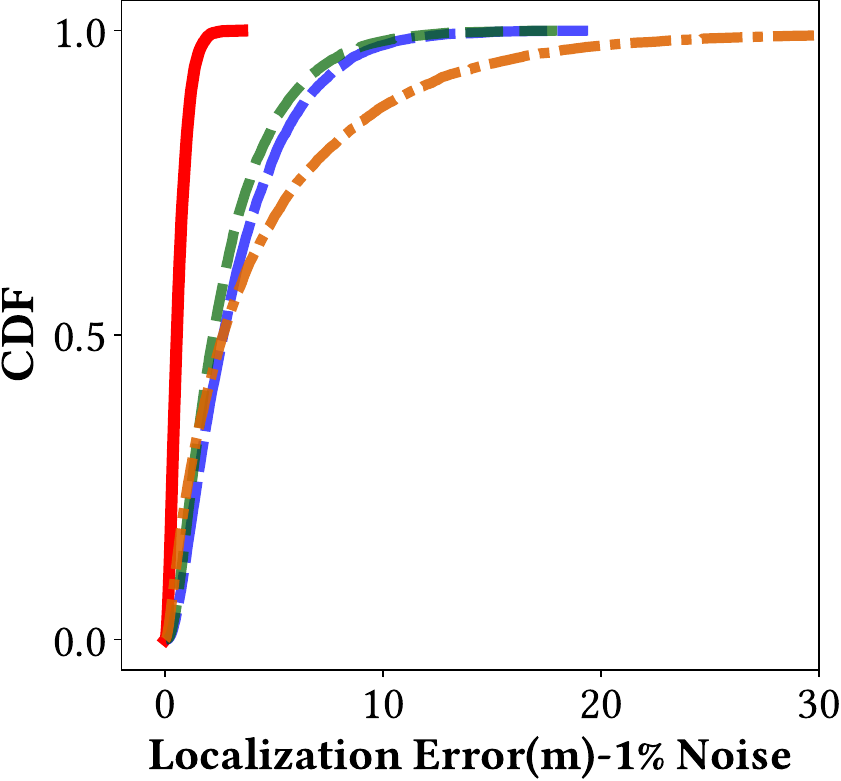}
      \hfill
      \includegraphics[height= 0.5\textwidth]{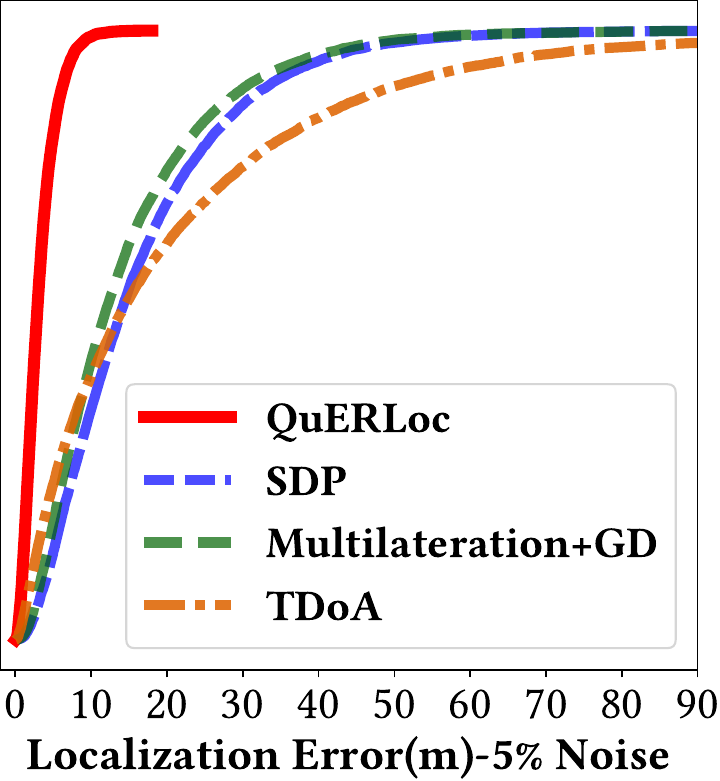}
      \label{fig:cdf_5}
      }
  \end{minipage}
  \caption{Performance of \sysname and baselines over different noise levels. {\rm  The CDF of localization error to all localization approaches when $m=3, 4$ and $5$ are plotted under noise levels $1\%$ and $5\%$.}}
  \label{fig:noisySim}
\end{figure*}

\subsubsection{Performance Metrics}
To evaluate the performance of a localization algorithm, we repeat the ranging and localization procedure under the same simulation settings, for a total of $r \in \mathbb{N}$ times. Denote $\hat{\bm{x}}^{(t)}$ and $\bm{x}^{(t)} := \bm{x}_{\text{real}}^{(t)}$ to be the estimation and ground truth of the sensor location in the $t^{\text{th}}$ iteration. To measure the precision of all presented localization techniques, we examine both localization error $\left\|\hat{\bm{x}}^{(t)} - \bm{x}^{(t)}\right\|$ of a single experiment and Root-Mean-Square-Error (RMSE) \cite{Ground_Based_Loc} of $r$ iterative experiments at the same noise level. Specifically, the RMSE is defined by
$$
    \text{RMSE} := \sqrt{\frac{1}{r} \sum_{t=1}^r \left\|\hat{\bm{x}}^{(t)} - \bm{x}^{(t)} \right\|^2 }.
$$

\subsubsection{Cram\'er-Rao Lower Bound}
Additionally, we examine the Cramér-Rao Lower Bound (CRLB) as a benchmark to gauge the optimal accuracy attainable by the estimator $\hat{\bm{x}}$ employed by \sysname. Recall that we derive the optimization objective in \eqref{eqn:likelihood} through the following log-probability density function:
$$
\begin{aligned}
    \log \mathcal{L} \left( \widetilde{\bm{\lambda}}^{(t)}   \right) &= - \sum_{k=1}^m \log \left( \sqrt{2\pi} \rho \widetilde{\lambda}_k^{(t)} \right) \\
    &- \frac{1}{2\rho^2} \left\| \sqrt{\widetilde{W}^{(t)}}\left(\bm{L}^{(t)} \bm{x} - \widetilde{\bm{h}}^{(t)}\right)\right\|^2.
\end{aligned}
$$
The \textit{Fisher information matrix} \cite{FIM} of the log-likelihood function can be formulated as 
\begin{equation}
    \label{eqn:FIM}
    \begin{aligned}
    \mathcal{F}\left(\widetilde{\bm{\lambda}}^{(t)}\right) &= -\mathbb{E} \left[ \frac{\partial^2 \log \mathcal{L} ( \widetilde{\bm{\lambda}}^{(t)}) }{\partial \bm{x} \partial \bm{x}^T} \right] = \left\{ -\mathbb{E} \left[ \frac{\partial^2 \log \mathcal{L} ( \widetilde{\bm{\lambda}}^{(t)} ) }{\partial x_i \partial x_j} \right] \right\}_{ij} \\
    &= \left\{  \left(\bm{L}^{(t)} \frac{\partial^2 \bm{x}}{\partial x_i \partial x_j } \right)^T \frac{1}{\rho^2} \mathbb{E} \left[ \widetilde{W}^{(t)}  \left(\bm{L}^{(t)} \bm{x} - \widetilde{\bm{h}}^{(t)}\right)\right]\right. \\
    &\left.+ \frac{1}{\rho^2} \left(\bm{L}^{(t)} \frac{\partial \bm{x}}{\partial x_i}  \right)^T \mathbb{E}\left[\widetilde{W}^{(t)}\right] \left(\bm{L}^{(t)} \frac{\partial \bm{x}}{\partial x_j} \right)    \right\}_{ij} \\
    &\approx \frac{1 + 3\rho^2}{\rho^2} \left(\bm{L}^{(t)}\right)^T \widetilde{W}^{(t)} \bm{L}^{(t)}.
    \end{aligned}
\end{equation}
The first term in the summation vanishes due to $\frac{\partial^2 \bm{x}}{\partial x_i \partial x_j} = \left[\frac{\partial \delta_{kj}}{\partial x_i}\right]_k =  \bm{0}$. The final expression in \eqref{eqn:FIM} originates from $\partial \bm{x}^T / \partial \bm{x} = \bm{I}$, and $\mathbb{E}[\widetilde{\lambda}_k^{-2}] = \lambda^{-2}_k \cdot \mathbb{E}[1-2\delta_k + 3\delta_k^2 + O(\delta_k^3)] \approx (1 + 3\rho^2) \lambda_k^{-2}$ by Taylor expansion along with $\mathbb{E}[\delta_k] = 0$ and $\mathbb{E}[\delta_k^2] = \text{Var}[\delta_k] + \mathbb{E}^2[\delta_k] = \rho^2$. CRLB provides a lowerbound $\mathbb{E}[(\hat{\bm{x}}^{(t)} - \bm{x}^{(t)} ) (\hat{\bm{x}}^{(t)} - \bm{x}^{(t)} )^T ] \succeq \mathcal{F}^{-1}(\widetilde{\bm{\lambda}}^{(t)})$. This allows us to derive a lowerbound for RMSE when $r$ is sufficiently large, 
\begin{equation}
\begin{aligned}
    \text{RMSE} &\overset{r \gg 1}{=} \sqrt{\frac{1}{r} \sum_{t=1}^r \mathbb{E}\left[ \left(\hat{\bm{x}}^{(t)} - \bm{x}^{(t)} \right)^T \left(\hat{\bm{x}}^{(t)} - \bm{x}^{(t)} \right) \right]} \\
    &= \sqrt{ \frac{1}{r} \sum_{t=1}^r \text{Tr} \left( \mathbb{E}\left[ \left(\hat{\bm{x}}^{(t)} - \bm{x}^{(t)} \right) \left(\hat{\bm{x}}^{(t)} - \bm{x}^{(t)} \right)^T \right] \right)} \\
    &\overset{\text{CRLB}}{\geq} \sqrt{\frac{1}{r} \sum_{t=1}^r \text{Tr} \left(\mathcal{F}^{-1}(\widetilde{\bm{\lambda}}^{(t)} )\right)  }.
\end{aligned}
\end{equation}

\subsection{Evaluation Results}
We evaluate all the positioning instances on a classical computer. Within the same setup, we repeatedly generate $r = 10^4$ locations and perturb the distances data in an analogous way to \eqref{eqn:noise_form}. 
\par All approaches including \sysname and baselines \textbf{\textit{(i)-(iii)}} will share identical testbed settings and estimate the same randomly generated sensor locations $\{\bm{x}^{(t)} \}_{t=1}^r$. The choices of used anchors are determined by the particular protocols of each localization method according to their selective strategies. Notice that we mainly focus on \sysname's capability to acquire special distance combinations rather than the enhancement quantum metrology would bring to the readout precisions \cite{AdvQuantumMetrology}. Thus, we set the factor $\rho$ to be identical among all approaches adopted at the same noise level.
\par We implemented all algorithms in Python, where the least-square regressions were solved using the built-in Python package numpy, and SOCPs/SDPs were solved using MOSEK \cite{mosek}. All simulations were run on a Windows PC with 16GB memory and AMD Ryzen 9 7945HX CPU.

\subsubsection{Performance with Few Numbers of Rangings.}
For each choice of the number of rangings $m \in \{3, 4, 5\}$, we set the probe scheme to be $I_k = \{ 2k-1, 2k  \} \subset \{1, \dots, n\}$ and $w_{i, k} = -(-1)^{i}$ for $1 \leq k \leq m$ and $i \in I_k$. {\sysname exploits $m$ distinct pairs of anchors for one-shot localization. Baselines approaches, including TDoA (where one anchor serves as the reference node across all TDoA measurements), utilize $m$ anchors since each ranging only introduces information from a single new anchor}. \fig \ref{fig:rmse_3}, \ref{fig:rmse_4} and \ref{fig:rmse_5} report the RMSE of localization approaches with respect to the varying noise factor when $m = 3, 4$, and $5$. \sysname works nicely under all presented cases, and consistently surpasses all the baseline methods. From the cumulative distribution function (CDF) of localization error in corresponding experiments under noise level $1\%$ and $5\%$ reported by \fig \ref{fig:cdf_3}, \ref{fig:cdf_4} and \ref{fig:cdf_5}, we observe that \sysname achieves high localization accuracy for the majority of test cases, producing comparatively few estimation of significant deviations. It is noteworthy that when few ($=3$) numbers of rangings are available in the $3$-dimensional space, \sysname can still produce satisfactory location estimation and closely follows the CRLB, while the outputs of all proposed baselines yield large deviation from the ground truth. The reason is that with the aid of \rangname, the objective set of optimization problem would degenerate from the intersection of a collection of curved surfaces to that of a collection of hyperplanes.

\begin{figure*}[t]
  \begin{minipage}{0.315\linewidth}
    \centering
    \subfloat[RMSE, $m = 3$]{
    \includegraphics[width=\textwidth]{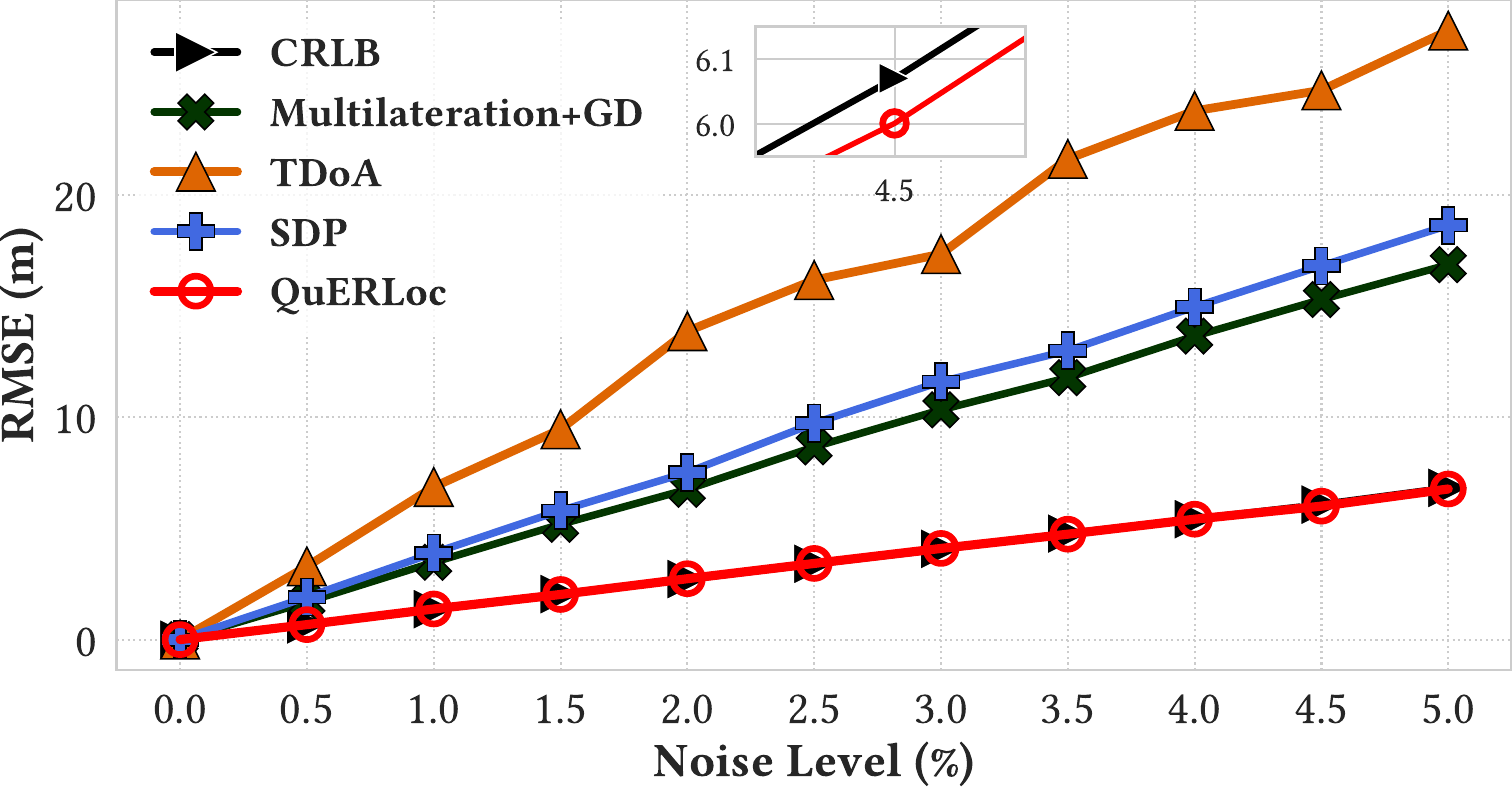}
    \label{fig:rmse_dup_3}
    }
  \end{minipage}
  \hfill
  \begin{minipage}{0.315\linewidth}
    \centering
    \subfloat[RMSE, $m=4$]{
        \includegraphics[width=1.0\textwidth]{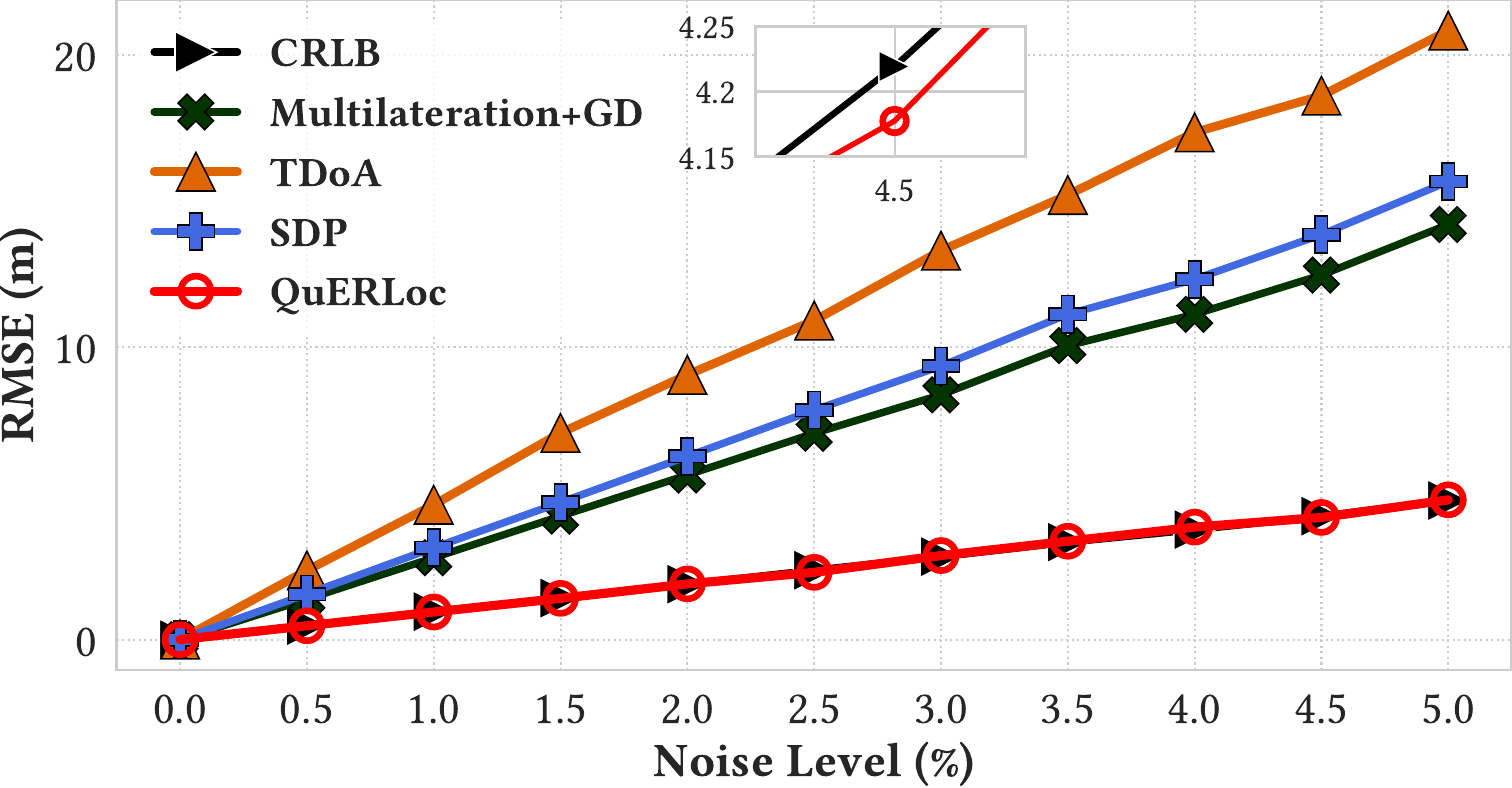}
        \label{fig:rmse_dup_4}
    }
  \end{minipage}  
  \hfill 
  \begin{minipage}{0.315\linewidth}
    \centering
    \subfloat[RMSE, $m=5$]{
    \includegraphics[width=\textwidth]{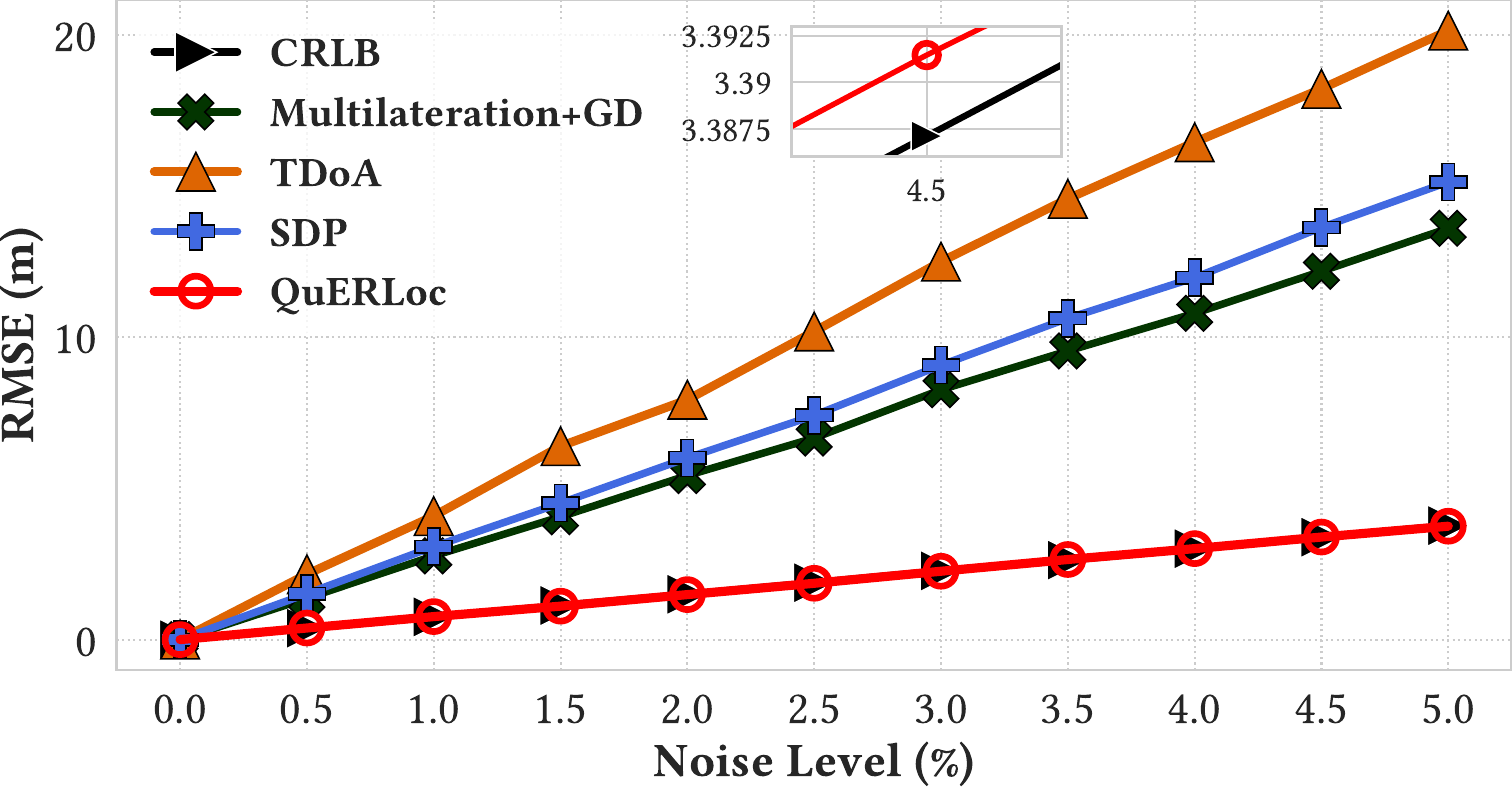}
    \label{fig:rmse_dup_5}
    }
  \end{minipage}
  \begin{minipage}{0.315\linewidth}
      \centering
      \subfloat[CDF, $m=3$]{
      \includegraphics[height= 0.5\textwidth]{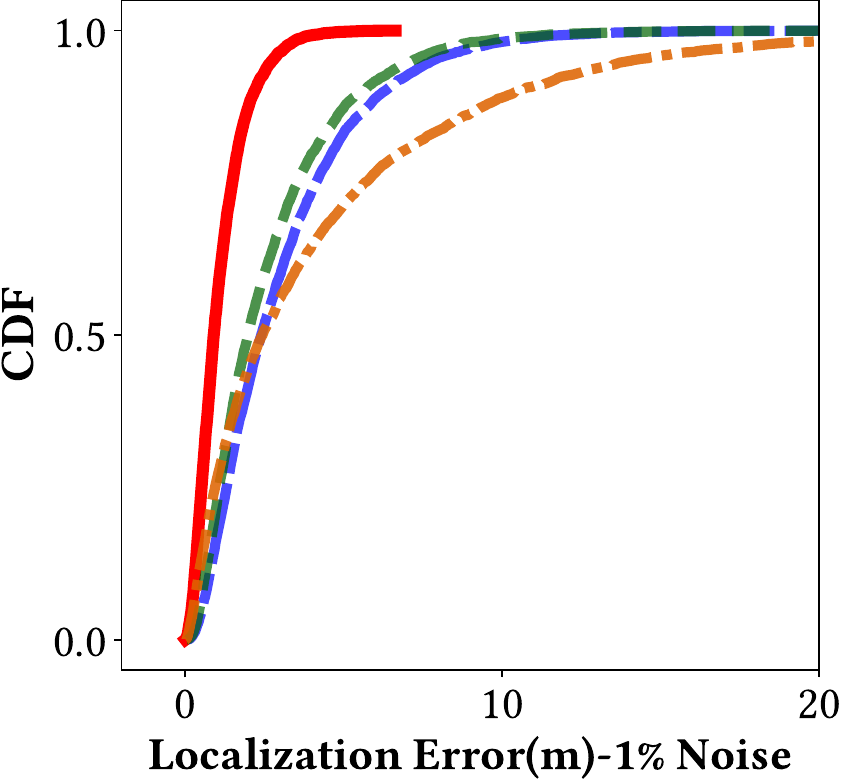}
      \hfill
      \includegraphics[height= 0.5\textwidth]{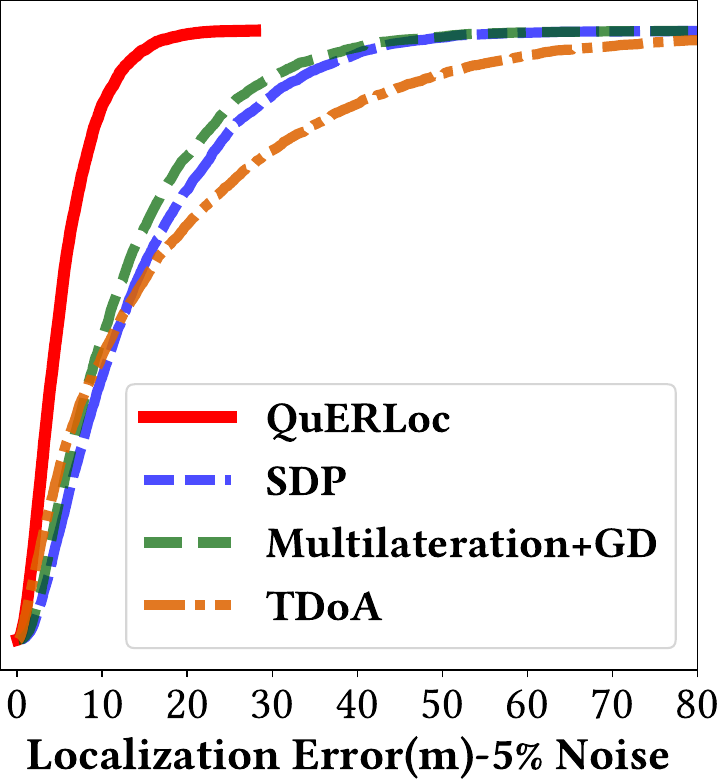}
      \label{fig:cdf_dup_3}
      }
  \end{minipage}
  \hfill 
  \begin{minipage}{0.315\linewidth}
      \centering
      \subfloat[CDF, $m=4$]{
      \includegraphics[height= 0.5\textwidth]{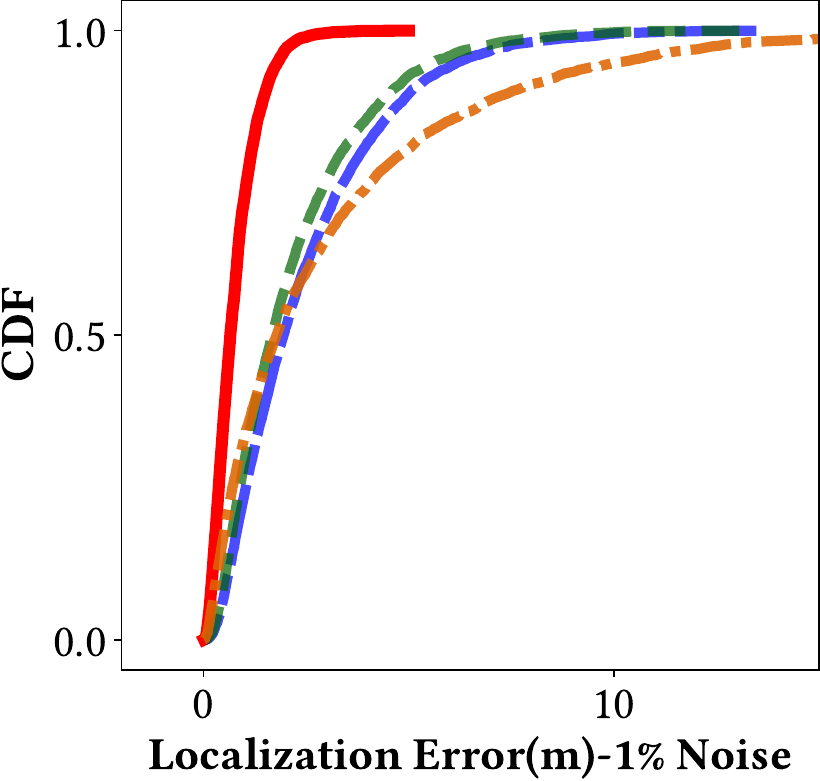}
      \hfill
      \includegraphics[height= 0.5\textwidth]{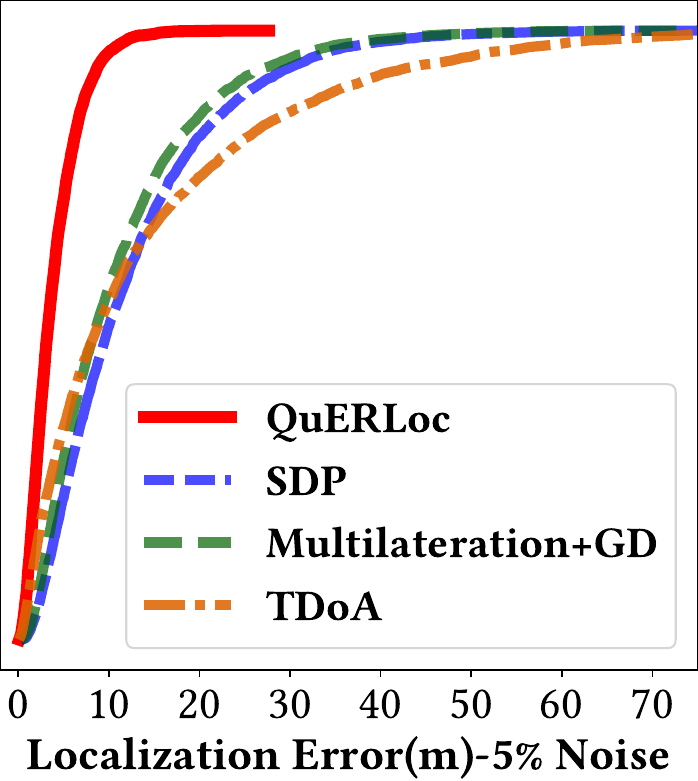}
      \label{fig:cdf_dup_4}
      }
  \end{minipage}
  \hfill 
  \begin{minipage}{0.315\linewidth}
      \centering
      \subfloat[CDF, $m=5$]{
      \includegraphics[height= 0.5\textwidth]{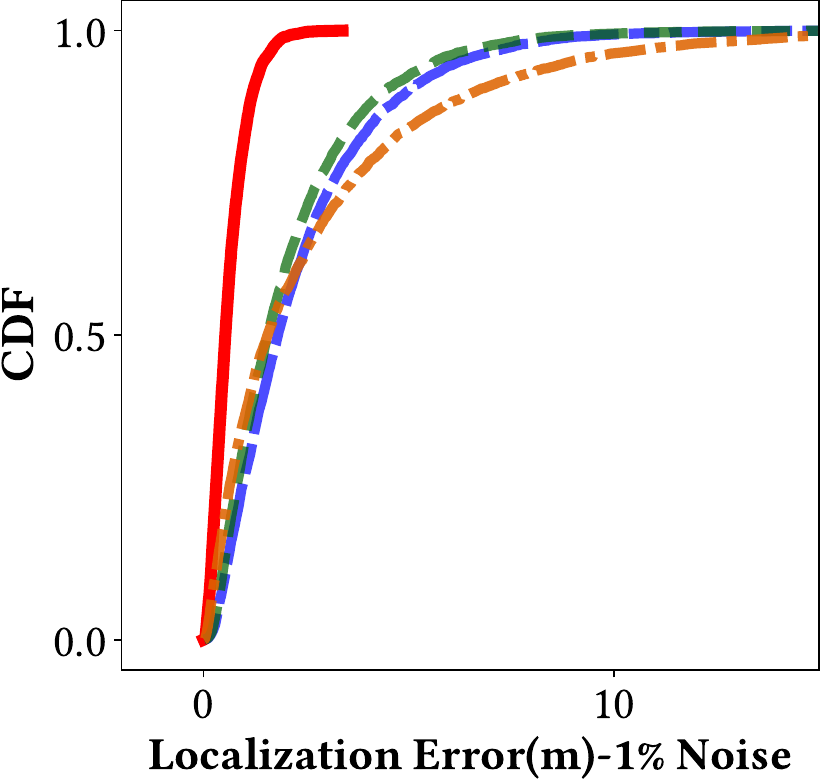}
      \hfill
      \includegraphics[height= 0.5\textwidth]{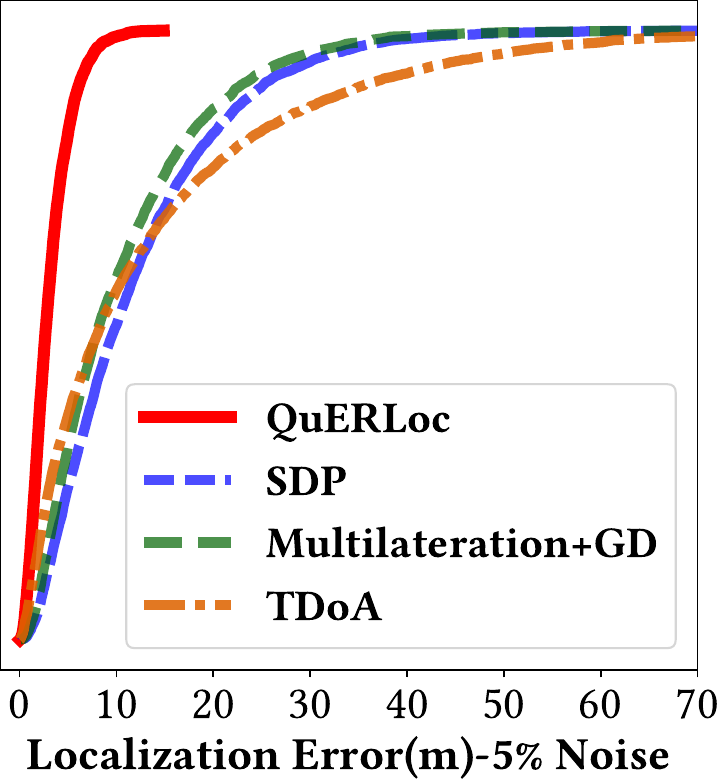}
      \label{fig:cdf_dup_5}
      }
  \end{minipage}
  \caption{Performance of \sysname and baselines with same anchor utilization. {\rm \sysname and all baseline methods conduct distance ranging with the same set of anchors.}}
  \label{fig:sameAnchors}
\end{figure*}

\subsubsection{Superiority of \sysname with Same Anchor Utilization}
One may doubt that the superiority of \sysname merely originates from the full utilization of available anchors, as in the previous experiment, \sysname used twice as many anchor nodes as baselines. 
We address this question by doubling the quantity of distance ranging for baselines (\ie, they are conducting $2m$ rangings using the same anchors utilized by \sysname) while maintaining that of \sysname at $m$. As shown in \fig \ref{fig:sameAnchors}, despite noticeable improvement in the performance of baselines, \sysname still largely outperforms them. As $m$ changes from $4$ to $5$, baselines only yield marginal accuracy improvement. \sysname achieves an RMSE of 27\% compared to the best-performing baseline Multilateration + GD, as shown in \fig \ref{fig:rmse_dup_5}.

\subsubsection{Case Study: Mimicking \sysname with Classical Ranging} 

One question might be raised naturally: Given that a distance combination analogous to \eqref{eqn:signal_form} is central to the superiority of \sysname, is it possible to mimic such a ranging process with classical ranging, thus achieving the same localization accuracy? We explore the feasibility of this approach by conducting the following experiment: For each instance of \rangname, classical ranging on each involved target-anchor distance $d_i$ is meanwhile conducted and combined. As for the systematic noise, we perturb corresponding readouts for \sysname and classical simulating system, denoted as {$\sysname\text{-sim}$}, as follows:
$$
\begin{aligned}
    &\sysname: \quad \widetilde{\lambda}_k = \left(1 + \delta_k\right) \cdot \sum_{i \in I_k} w_{i, k} d_i^2, \quad \delta_k \sim \mathcal{N}(0, \rho).  \\
    &\sysname\text{-sim}: \quad \widetilde{\lambda}_k' = \sum_{i \in I_k} w_{i, k} \left[ d_i \cdot \left(1 + \delta_i\right)  \right]^2, \quad \delta_i \sim \mathcal{N}(0, \rho).
\end{aligned}
$$
\begin{figure}[htbp]
    \centering
    \includegraphics[width=0.83\linewidth]{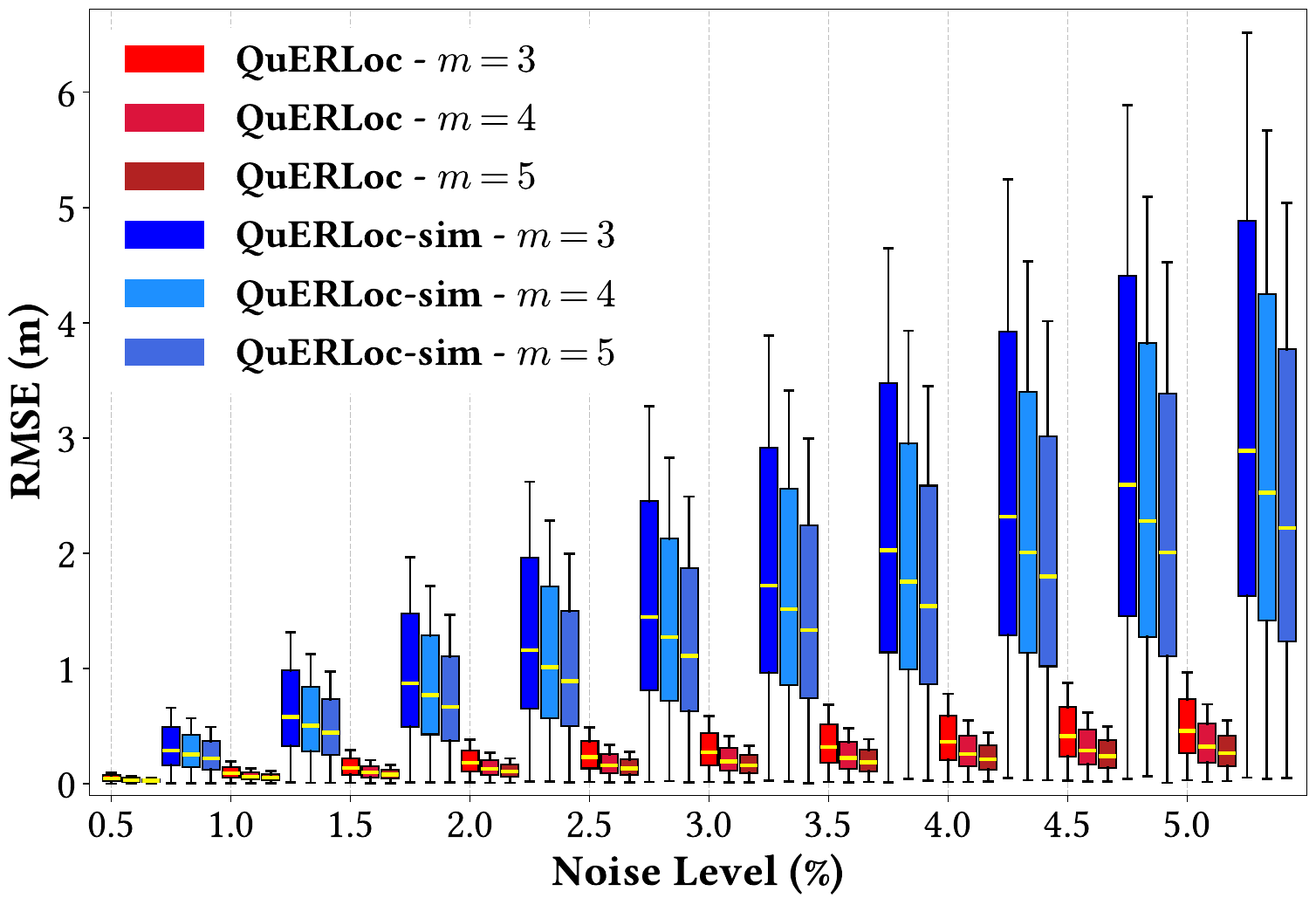}
    \caption{Comparison of \sysname and \sysname-sim with $m = 3, 4, 5$ under non-zero noise levels $0.5\%$ to $5\%$. {\rm Each cluster of boxes corresponds to one noise level.}}
    \label{fig:classical_querloc}
\end{figure}

The localization performance under both settings is compared in \fig \ref{fig:classical_querloc}. It be observed that the \sysname-sim suffers from evident deterioration in performance, as it doubles the quantity of ranging required compared to \rangname under the designed probing scheme, and the noise is integrated into the system in a quadratic manner.

\subsubsection{Time Complexity}
\label{sec:complexity}
Choosing localization methods involves a trade-off between accuracy and latency. Trilateration-based localization requires low running time but is highly sensible to noise, while conic relaxation and gradient-descent methods have no guarantee of instantaneous localization. As is illustrated in \tab \ref{tab:complexity} the computational complexity and time consumption of several localization approaches when $\rho=5\%$, \sysname provides reliable localization with much more efficient computational requirements. On average, QuERLoc consumes only 2.4\% of the time required by the most efficient baseline method Multilateration + GD.

\begin{table*}[htbp]
\begin{tabular}{lll}
\hline
\textbf{Localization Methods}            & \textbf{Time Complexity }                                                                       & \textbf{Average Time Consumption (sec)} \\ \hline
$\sysname$         & ${O}\left(d^2(m+d)\right)$                                                & $3.27 \times 10^{-4}$        \\
SDP                &${O}\left( \sqrt{d}(md^2 + m^\omega + d^\omega) \log(1/\alpha) \right)$ ~~\# state-of-the-art \cite{SOTA_SDP_Complexity}~ & $2.38\times 10^{-2}$         \\
TDoA               & ${O}\left(d^2(m+d) + T_{{SOCP}}\right)$                              & $2.03\times 10^{-2}$         \\
Multilateration+GD & ${O}\left(d^2(m+d) + M d^2 \right) $                                      & $1.37 \times 10^{-2}$        \\ \hline
\end{tabular}
\caption{Complexity and latency, $\rho=5\%$. {\rm $\alpha > 0$ is the relative accuracy, $\omega$ is the exponent of matrix multiplication, and $M$ is the number of BFGS (default choice of gradient-descent search in cvxpy \cite{cvxpy}) iterations. Asymptotic behaviour of $T_{{SOCP}}$ depends on the solver's adaptive choice of problem reduction into various conic programming instances. When reduced to SDP, $T_{{SOCP}} = {O}( \sqrt{d}(md^2 + m^\omega + d^\omega) \log(1/\alpha))$.}}
\label{tab:complexity}
\end{table*}

\section{Conclusion}
\label{sec:conclusion}
In this paper, we present \sysname, a novel localization approach that exploits the advantage of quantum-enhanced ranging realized by quantum metrology with entangled probes. We propose a new distance ranging model based on the quantum control theory and phase estimation by fine-tuning dynamics of quantum probes under two-level approximation, which we call \rangname. 
We show that by a specially designed probe state, quantum-enhanced ranging can result in a convex optimization problem, which can be solved efficiently.
Extensive simulations verify that \sysname significantly outperforms baseline approaches using classical ranging and saturates CRLB, demonstrating its superiority in both accuracy and latency. Our work provides a theoretical foundation for a potential application of quantum metrology in the field of range-based localization. We believe \sysname leads the research on localization with quantum resource and opens new directions to both fields of quantum computing and mobile computing.

\bibliographystyle{ACM-Reference-Format}
\bibliography{refs_mobihoc}

\end{document}